\documentstyle[aps]{revtex}

\newcommand{\r}[2]{\mbox{$\langle \hat{c}_{#1}^{\dag} \hat{c}_{#2} \rangle$}}

\newcommand{\k}[2]{\mbox{$\langle \hat{c}_{#1} \hat{c}_{#2} \rangle$}}
\newcommand{\ck}[2]{\mbox{$\langle \hat{c}_{#1}^{\dag} \hat{c}_{#2}^{\dag} \rangle$}}

\begin{document}

\title{Basis-dependent dynamics of trapped Bose-Einstein condensates and analogies with semi-classical laser theory}
\author{N.P. Proukakis$^{1,3}$ and P. Lambropoulos$^{1,2}.$}
\address{$^{1}$Foundation for Research and Technology Hellas, Institute of Electronic Structure and Laser, P.O. Box 1527, Heraklion 71 110, Crete, Greece.}
\address{$^{2}$Department of Physics, University of Crete, Heraklion 71110, Crete, Greece. }
\vspace{6mm}
\maketitle


\begin{abstract}

We present a consistent second order perturbation theory for the lowest-lying condensed modes of very small, weakly-interacting Bose-Einstein condensates in terms of bare particle eigenstates in a harmonic trap. After presenting our general approach, we focus on explicit expressions for a simple three-level system, mainly in order to discuss the analogy of a single condensate occupying two modes of a trap with the semi-classical theory for two-mode photon lasers. A subsequent renormalization of the single-particle energies to include the dressing imposed by mean fields demonstrates clearly the consistency of our treatment with other kinetic approaches. 

\end{abstract}

pacs: 03.75.Fi, 42.55.Ah, 03.75.-b, 42.50.-p

\section{Introduction}

The kinetic theories of Bose-Einstein condensation in a trap can be divided into two general categories. The fully quantum approaches have been based either on a perturbative master equation treatment for the many-body density matrix \cite{QKV}, or on a single Fokker-Planck equation for the nonequilibrium dynamics of the entire system \cite{Stoof_FP}. Alternative approaches to the nonequilibrium dynamics of trapped Bose-condensed gases are essentially of a perturbative nature, based on a suitably truncated coupled equation of motion hierarchy for normal and anomalous averages \cite{Prouk_NIST,Prouk_T_Matrix,Walser_QK,Walser_Sim,Griffin_Hydro,Milena,MF}. The growth of condensation was first studied numerically by Gardiner and co-workers \cite{QKVI,QKVII}, based on the theory developed in \cite{QKV}, and their findings were in qualitative agreement with the description of Kagan, Svistunov and Shlyapnikov \cite{Kagan}. Independent studies by Stoof and co-workers, based on similar approximations, have produced growth curves \cite{Stoof_Growth} which are in very good agreement with those of Gardiner et al. \cite{QKVII}. 

The large condensates typically produced in experiments lead to large mean field potentials which significantly dress the single-particle eigenenergies of the trap potential, thus making it convenient to work in a suitably dressed basis. The conventional theoretical picture adopted is thus that of a very large single-mode condensate interacting with a large number of quasiparticles and higher-lying excited atoms, whereas it is not a priori necessary that single-mode condensation will correspond to all experimental conditions \cite{Stoof_FP,Stoof_Private}. Instead of working in the usual `condensed matter' approach of a condensate and a set of quasiparticle excitations which are, by definition, orthogonal to it (which is suitable for large condensates) we focus our description in terms of a more `quantum-optical' approach, i.e. in terms of individual modes of the trapping potential \cite{Prouk_NIST,Prouk_T_Matrix,Walser_QK,Walser_Sim}, for which one must in principle consider the condensate spanning a large number of modes, with the same modes being simultaneously occupied by non-condensate atoms. Within such a picture, this paper explicitly discusses the simplest deviation from single-mode condensation in the context of a bare single-particle basis, in a manner analogous to existing treatments. Formulating the problem in terms of bare particle eigenenergies can only be useful in the case of extremely small, dilute, weakly-interacting condensates, when the mean field effects are small enough that their induced shifts of the single-particle energies can be treated as perturbations. Such an approach is nevertheless beneficial for discussing the inherent multi-mode nature of trapped condensates and their relation to the corresponding (semi-classical) theory of multi-mode photon lasers.

The first part of this paper (Sec. II) reviews in the usual manner the formal development of the coupled equation of motion approach. To simplify the physical picture and bring out the underlying structure in a clear manner, we focus on a very simple system consisting of the three lowest trap eigenstates, for which we derive explicit equations of motion in the Popov approximation \cite{Popov} in Sec. III (with corresponding off-equilibrium contributions given in Appendix A, and some further clarifications in Appendix B). By explicitly discussing the interplay of two coupled condensed modes, we show how our treatment reduces to the Hartree-Fock theory for binary condensates \cite{HF_Theory} (Sec. IV A1). An important advantage of formulating the theory in terms of bare single-particle eigenstates is that it allows us to discuss in detail the analogy between our equations for the coupled dynamics of two condensed modes, and the corresponding ones arising in the semi-classical treatments of two-mode photon lasers (Sec. IV A2). We show that, the extent to which this analogy can be drawn for the inherent multi-mode nature of a single condensate, depends critically on the choice of the single-particle basis  (i.e. bare or dressed by various types of mean fields) in terms of which the analysis is carried out. Our discussion here is distinct from conventional analogies  based on two different condensates which are spatially separated \cite{Menotti}, in different spin states  \cite{Atom_Laser_Review}, or outcoupled by radiation applied at two different frequencies \cite{2Mode}. Sec. IV B compares our approach to conventional kinetic treatments, where we show explicitly that our theory reduces to the multi-mode kinetic treatment of Walser et al., upon shifting our description to a basis in which the single-particle eigenenergies become dressed by the usual Hartree-Fock-Bogoliubov (HFB) mean field potentials. This suggests that, contrary to an implication in \cite{Walser_QK}, such dynamic equations (and collisional integrals) are inevitably basis-dependent.

\section{The Coupled Equation of Motion Approach}

Consider a sufficiently dilute, weakly-interacting partially Bose-condensed trapped gas with a binary-interaction hamiltonian 
\begin{equation}
\hat{H} = \sum_{rs} \Xi_{rs}^{Bare}  \hat{a}_{r}^{\dag} \hat{a}_{s} + \frac{1}{2}\sum_{rsmn} V_{rsmn} \hat{a}_{r}^{\dag} \hat{a}_{s}^{\dag} \hat{a}_{m} \hat{a}_{n}
\end{equation}
Here  $\hat{\Xi}^{bare} = - (\hbar^{2} \nabla^{2})/(2m) + V_{trap}({\bf r})$ contains both kinetic energy and trapping potential and $V_{rsmn}$ represents the symmetrized form of the interaction potential between a pair of particles, defined by
$V_{rsmn}  = \frac{1}{2} \left\{ \langle rs| \hat{V} |  mn \rangle +   \langle rs| \hat{V} |  nm \rangle \right\}$, where $|i \rangle = \psi_{i}({\bf r})$ denotes a single-particle eigenstate of the trap.
The single-particle operators $\hat{a}_{i}$ are related to the Bose field operator $\hat{\Psi}({\bf r},t)$ via
$\hat{\Psi}({\bf r},t) = \sum_{i} \psi_{i}({\bf r}) \hat{a}_{i}(t)$.
We assume the system to be in a symmetry-broken phase and hence express the single-particle operators $\hat{a}_{i}(t)$ as \cite{Blaizot}
\begin{equation}
\hat{a}_{i}(t) = \langle \hat{a}_{i} \rangle + \left(\hat{a}_{i}- \langle \hat{a}_{i}  \rangle \right) = z_{i}(t) + \hat{c}_{i}(t)
\end{equation}
This allows, in general, for a coherent mean-field amplitude $z_{i}$ to form in a number of low-lying trap levels. We can now formulate a non-equilibrium theory for the coupled evolution of condensate mean field amplitudes and fluctuations about these values, based on a suitably truncated hierarchy of coupled equations of motion and appropriate decoupling approximations. 

Possibly the most direct approach for studying the dynamics in a closed system is based on solving an appropriate set of such equations self-consistently, in terms of  exact interatomic potentials; such an approach has been discussed, for example, in \cite{Prouk_NIST}. In a realistic system, the number of trap eigenstates will be very large, making such a procedure computationally very demanding. Since the most interesting dynamics take place in the low-lying levels, in this paper we have chosen to restrict our analysis to such levels (although such treatment will also implicitly yield the behavior of  high-lying thermal levels). Accepting a distinction between low- and high-lying levels allows us to  adiabatically eliminate all high-lying levels appearing as intermediate states in the equations for the  evolution of averages of low-lying states; this procedure is known to lead to the renormalization of the exact (single-vertex) interatomic potential to an effective two-body one (over high-lying states), as discused in \cite{Morgan}. We thus arrive at the situation where the hamiltonian of the system still has the general form of Eq. (1), but with the single-vertex interatomic potential $V$ replaced by an effective two-body T-matrix, $T$, over high-lying states, with the simultaneous restriction of all bare trap eigenstates being summed over low-lying levels. For sufficiently dilute systems at low temperature, this restricted effective two-body interaction is approximately equal to the full two-body T-matrix, $T^{2B}$, giving the scattering of two particles in vacuum. We thus approximate $T$  in terms of $T^{2B}$, by ensuring that purely two-body effects due to collisions occuring in vacuum are not double-counted (for a detailed discussion of the relation of these effective interactions and the renormalization required to avoid double-counting, the reader is referred to \cite{Morgan,Keith}). We note that it is precisely the quantity  $T^{2B}$  which corresponds, in three dimensions, to  the usual binary s-wave scattering length pseudopotential.

We are interested in working out in a self-consistent manner the evolution of condensate population and incoherent fluctuations about this value, and for consistency we work throughout this paper with averages of two single-particle operators. The Heisenberg equation of motion for such a general product of two operators is given by (setting $\hbar=1$)
\begin{eqnarray}
i  \frac{d}{dt}\langle   \hat{a}_{i}^{\dag} \hat{a}_{j} \rangle & &  =  \sum_{r}^{'} \left\{ \Xi_{jr}^{Bare}\langle \hat{a}_{i}^{\dag} \hat{a}_{r} \rangle -  \Xi_{ri}^{Bare} \langle \hat{a}_{r}^{\dag} \hat{a}_{j} \rangle \right\} \nonumber \\ & & + \sum_{rms}^{'} \left\{ T_{jsmr} \langle \hat{a}_{i}^{\dag} \hat{a}_{s}^{\dag}\hat{a}_{m}\hat{a}_{r} \rangle - T_{ismr}^{*} \langle \hat{a}_{m}^{\dag} \hat{a}_{r}^{\dag}\hat{a}_{s}\hat{a}_{j} \rangle \right\}
\end{eqnarray}
where the primes indicate summation over low-lying levels. Consistent application of second order perturbation theory in the weakly-interacting limit (i.e. when the system can be well described in terms of single-particle wavefunctions) should yield correct expressions for energy level shifts and population damping. To proceed with our treatment, we must determine whether there are any quantities (fluctuations) which evolve faster than others (mean fields), so that the former can be adiabatically eliminated from the first order expressions. For example, in the usual rate equation treatments, one eliminates all off-diagonal normal and anomalous averages in favour of (diagonal) populations; one often speaks of coherences damping out faster than populations due to the coupling of the system to its environment. This gives rise to a set of equations to second order in the effective potentials, coupling populations to populations, in what is often termed the secular approximation \cite{Cohen}.

One could, however,  argue that the choice of which low-lying averages can be adiabatically eliminated depends on the basis employed for the description of the system, i.e. essentially on whether the mean field energy shifts are correctly taken into account or not. Starting from bare trap eigenenergies, one would not expect any normal or anomalous averages to be slowly-evolving; hence, the correct second order expression of Eq. (3) in a bare particle basis can be obtained by adiabatic elimination of the entire quantity $\langle \hat{a}^{\dag} \hat{a}^{\dag} \hat{a} \hat{a} \rangle$ by means of its respective equation of motion. This is the procedure adopted in the main part of this work, which gives rise to complex equations appearing to contain additional terms when compared directly to similar treatments. However, a subsequent transformation to a dressed single-particle basis (i.e. dressed eigenenergies) shows clearly that such terms drop out from the respective equations of motion for populations in dressed eigenstates, as anticipated. Nonetheless, an approach in terms of bare single-particle eigenenergies allows one to draw important analogies between multi-mode condensation and multi-mode laser theory. Proceeding thus with the treatment in a bare basis, and taking the  operator $\hat{\Xi}^{bare}$ to be diagonal, we obtain
\begin{equation}
i  \frac{d}{dt}\langle \hat{a}_{i}^{\dag} \hat{a}_{s}^{\dag}\hat{a}_{m}\hat{a}_{r}  \rangle   =  (\omega_{m}+\omega_{r}-\omega_{i}-\omega_{s}) \langle \hat{a}_{i}^{\dag} \hat{a}_{s}^{\dag}\hat{a}_{m}\hat{a}_{r}  \rangle +F_{i}(T,\hat{a},\hat{a}^{\dag};t)
\end{equation}
where $\omega_{i}$ correspond to bare trap energies and $F_{i}(T,\hat{a},\hat{a}^{\dag};t) $ defines the collisional evolution of $\langle \hat{a}_{i}^{\dag} \hat{a}_{s}^{\dag}\hat{a}_{m}\hat{a}_{r}  \rangle$ in such a basis by \cite{Thesis}
\begin{eqnarray}
F_{i}(T,\hat{a},\hat{a}^{\dag})&  = & \sum_{lt}^{'} T_{mrlt} \langle \hat{a}_{i}^{\dag} \hat{a}_{s}^{\dag}\hat{a}_{l}\hat{a}_{t}  \rangle - \sum_{pq}^{'} T_{pqis} \langle \hat{a}_{p}^{\dag} \hat{a}_{q}^{\dag}\hat{a}_{m}\hat{a}_{r}  \rangle \nonumber \\
& + & \sum_{plt}^{'} T_{prlt} \langle \hat{a}_{p}^{\dag}\hat{a}_{i}^{\dag} \hat{a}_{s}^{\dag} \hat{a}_{m}\hat{a}_{l} \hat{a}_{t}  \rangle + \sum_{plt}^{'} T_{pmlt} \langle \hat{a}_{p}^{\dag} \hat{a}_{i}^{\dag} \hat{a}_{s}^{\dag} \hat{a}_{r} \hat{a}_{l} \hat{a}_{t}  \rangle \nonumber \\
& - & \sum_{pql}^{'} T_{pqls} \langle \hat{a}_{p}^{\dag} \hat{a}_{q}^{\dag} \hat{a}_{i}^{\dag} \hat{a}_{l} \hat{a}_{m} \hat{a}_{r}  \rangle - \sum_{pql}^{'} T_{pqli} \langle \hat{a}_{p}^{\dag} \hat{a}_{q}^{\dag} \hat{a}_{s}^{\dag} \hat{a}_{l} \hat{a}_{m} \hat{a}_{r}  \rangle \label{F}
\end{eqnarray}
Assuming real eigenvalues, we obtain the following exact integral relation
\begin{eqnarray}
\frac{d}{dt}\langle \hat{a}_{i}^{\dag}\hat{a}_{j} \rangle = & & -i \left(\omega_{j}-\omega_{i} \right) \langle \hat{a}_{i}^{\dag} \hat{a}_{j} \rangle -i \left[ \sum_{rms}^{'} T_{jsmr} \langle \hat{a}_{i}^{\dag} \hat{a}_{s}^{\dag} \hat{a}_{m} \hat{a}_{r} \rangle -e.c. \right] \nonumber \\
& & - \left\{ \sum_{rms}^{'} T_{jsmr} \int_{t_{0}}^{t} dt^{'}  e^{-i(\omega_{m} + \omega_{r} -\omega_{s}-\omega_{i})(t -
t^{'})} F_{i} ( T, \hat{a},\hat{a}^{\dag} ; t^{'}) + e.c. \right\} \label{FullPop}
\end{eqnarray}
where $e.c.$ stands for the exchange conjugate (i.e. conjugate expression with labels i and j interchanged). By using the definition $\hat{a}_{i} = z_{i} + \hat{c}_{i}$, we obtain all second-order collisional terms of our approach. 
However, for these to be useful, we must express them in terms of a closed system of equations, by imposing suitable  approximations.
Firstly we  decouple averages containing more than two single-particle operators via
\begin{equation}
\langle \hat{c}_{r}^{\dag} \hat{c}_{s}^{\dag} \hat{c}_{m} \hat{c}_{n} \rangle \approx \langle \hat{c}_{r}^{\dag}  \hat{c}_{m}  \rangle \langle \hat{c}_{s}^{\dag}  \hat{c}_{n} \rangle +\langle \hat{c}_{r}^{\dag}  \hat{c}_{n}  \rangle \langle \hat{c}_{s}^{\dag}  \hat{c}_{m} \rangle  +    \langle \hat{c}_{r}^{\dag}  \hat{c}_{s}^{\dag}  \rangle \langle \hat{c}_{m}  \hat{c}_{n} \rangle         \label{Four2}
\end{equation}
and
\begin{eqnarray}
\langle \hat{c}_{p}^{\dag} \hat{c}_{r}^{\dag} \hat{c} _{s}^{\dag} \hat{c} _{q} \hat{c}_{l} \hat{c}_{t} \rangle \approx & & \langle \hat{c}_{p}^{\dag} \hat{c}_{q} \rangle \left( \langle \hat{c}_{r}^{\dag} \hat{c}_{l} \rangle  \langle \hat{c}_{s}^{\dag} \hat{c}_{t} \rangle + \langle \hat{c}_{r}^{\dag} \hat{c}_{t} \rangle  \langle \hat{c}_{s}^{\dag} \hat{c}_{l} \rangle \right) \nonumber \\
& + & \langle \hat{c}_{p}^{\dag} \hat{c}_{l} \rangle \left( \langle \hat{c}_{r}^{\dag} \hat{c}_{q} \rangle  \langle \hat{c}_{s}^{\dag} \hat{c}_{t} \rangle + \langle \hat{c}_{r}^{\dag} \hat{c}_{t} \rangle  \langle \hat{c}_{s}^{\dag} \hat{c}_{q} \rangle \nonumber \right) + \langle \hat{c}_{p}^{\dag} \hat{c}_{t} \rangle \left( \langle \hat{c}_{r}^{\dag} \hat{c}_{q} \rangle  \langle \hat{c}_{s}^{\dag} \hat{c}_{l} \rangle + \langle \hat{c}_{r}^{\dag} \hat{c}_{l} \rangle  \langle \hat{c}_{s}^{\dag} \hat{c}_{q} \rangle \right) \nonumber \\
& + & \k{q}{l} \left( \ck{p}{r} \r{s}{t} + \ck{p}{s} \r{r}{t} + \ck{r}{s} \r{p}{t} \right) \nonumber \\ 
& + & \k{q}{t} \left( \ck{p}{r} \r{s}{l} + \ck{p}{s} \r{r}{l} + \ck{r}{s} \r{p}{l} \right) \nonumber \\
& + &  \k{l}{t} \left( \ck{p}{r} \r{s}{q} + \ck{p}{s} \r{r}{q} + \ck{r}{s} \r{p}{q} \right) \label{Six2}
\end{eqnarray}
.

Based on our formulation in terms of effective interactions, we further impose the Markov approximation which takes the quantity $F_{i}(T,\hat{a},\hat{a}^{\dag})$ out of the integrand and defines the intermediate propagators. This implies the assumption that the operators $\hat{a}_{i}$ evolve freely between collisions via $\hat{a}_{i}(t^{'})=e^{+i \omega_{i} (t-t^{'})} \hat{a}_{i}(t) $. Thus, the second order contributions of  Eq. (\ref{FullPop}) acquire the general form
\begin{equation}
\left[ \frac{d}{dt} \langle \hat{a}_{i}^{\dag}\hat{a}_{i} \rangle  \right]_{T^{2}} = -\sum_{rms}^{'} \left\{ \left[ T_{ismr} \sum_{\cdots} \left( \int_{t_{0}}^{t} d\tau  e^{-i(\Delta \omega_{\cdots}) (t-t^{'})} \right) T_{\cdots} \tilde{F_{i \cdots}}(t) \right] + e.c. \right\} \label{FullPop2}
\end{equation}
where the above quantity $ T_{\cdots} \tilde{F_{i \cdots}}(t) $ corresponds to $F_{i}$ defined by Eq. (5), and this notation has been used to indicate that the dotted indices of the second $T$ are the {\em same} indices as the ones appearing in the exponential of the integrand, as a result of the Markov approximation. Following the notation of Gardiner et al. \cite{QKV}, we now re-write the above integral as
\begin{equation}
\lim_{\eta \rightarrow 0^{+}} \int_{0}^{\infty} d \tau e^{ \pm i (\Delta \omega \pm i \eta) \tau} =  \delta^{(p)}(\Delta \omega) =   \pi \delta(\Delta \omega) \pm i {\cal P} \left( \frac{1}{ \Delta \omega} \right)
\end{equation}
where ${\cal P} \left( \frac{1}{ \Delta \omega} \right)$ corresponds to the principal value of the integral and the upper limit of integration has been approximated by $(t-t_{0}) \rightarrow \infty$, based on the usual assumption that successive collisional events are well separated in time. Having discussed our approximations,  we can now write down the coupled rate equations for coherent, incoherent and total populations.

\section{Rate Equations for Multi-level Condensation}

In this section, we focus on the application of the above methodology to a simple three-level system which we discuss within the Popov approximation \cite{Popov}, in which anomalous averages of the non-condensate  are ignored in the final expressions. We are well aware that this system is rather idealized and do not claim that it will accurately reproduce the entire dynamics of condensed and thermal atoms. In fact, corrections beyond Popov may be significant, as shown in this context by Walser et al. \cite{Walser_Sim} (see also \cite{Prouk_1D,Hutch_PRL,Prouk_JPhysB}), and we defer their explicit discussion to a subsequent paper (but see also Sec. IV. B). In the current paper we are mainly concerned with addressing the simplest departure from single-level condensation, and how this might affect the dynamics of low-lying levels. To this aim, such a system is ideal for a simple comparison with other kinetic theories \cite{QKV,Stoof_FP,Walser_QK,Griffin_Hydro,Milena,Stoof_Growth}. Perhaps more significantly, such a small system will enable us to discuss in an explicit manner the desired analogy of the multi-mode nature of a single condensate with multi-mode semiclassical laser theory \cite{Mandel}.  Our notation is as follows: The total population $N_{i}$ in level $i$ is written as a sum of coherent $|z_{i}|^{2}$ and incoherent $n_{i}=\rho_{ii}$ populations, via
\begin{equation}
N_{i} = \langle \hat{a}_{i}^{\dag} \hat{a}_{i} \rangle =\langle \hat{c} _{i}^{\dag}\hat{c}_{i} \rangle + |z_{i}|^{2} =  n_{i} + |z_{i}|^{2}
\end{equation}
with off-diagonal coherent and incoherent matrix elements respectively defined by $\zeta_{ij} = \left( z_{j}^{*} z_{i} \right)$ and $\rho_{ij} = \langle \hat{c}_{j}^{\dag} \hat{c}_{i} \rangle$.

\subsection{Evolution of Total Populations}

We start with expressions for the coupled evolution of total populations in the three lowest-lying bare levels under the assumption that all of them may exhibit partial condensation. Since we are dealing with a closed system, the evolution of populations in these levels trivially satisfy
\begin{equation}
\frac{dN_{0}}{dt}=\frac{dN_{2}}{dt}=-\frac{1}{2} \left( \frac{dN_{1}}{dt} \right)
\end{equation}
where
\begin{equation}
\frac{dN_{0}}{dt}= -2i \left[ T_{0211} \left(\rho_{10}+\zeta_{10} \right) \rho_{12} -c.c. \right] + \left\{ \left( R_{T_{|0211|^{2}}} + \tilde{Q}_{ijji}^{0211} \right) + c.c. \right\}
\end{equation}
Here we have defined the collisional redistribution rate
\begin{eqnarray}
R_{T_{|0211|^{2}}} = & &  2 \Gamma_{0211} \left[ (N_{0}+1)(N_{2}+1)N_{1}^{2} - N_{0}N_{2}(N_{1}+1)^{2} \right] \nonumber \\
& &- \Gamma_{0211}(1+N_{0}+N_{2})|z_{1}|^{4} + \tilde{R}_{T_{|0211|^{2}}} \label{R}
\end{eqnarray} 
where
\begin{equation}
\Gamma_{0211} =  |T_{0211}|^{2} \lim_{\eta \rightarrow 0^{+}} \int_{0}^{\infty} d \tau e^{ \pm i(\omega_{0}+\omega_{2}-2\omega_{1} \pm i \eta) \tau } = |T_{0211}|^{2} \delta^{(p)}(\omega_{0}+\omega_{2}-2\omega_{1})
\end{equation}
and the `tilde' in $ \tilde{R}_{T_{|0211|^{2}}} $ and $  \tilde{Q}_{ijji}^{0211}$  denotes contributions which are off-diagonal in  $\rho$ or $\zeta$, whose explicit  expressions can be found in Appendix A. In the context of the well-known Boltzmann scattering factors, the appearance of the additional term $ \sim \Gamma_{0211} |z_{1}|^{4}$ may appear somewhat perplexing. We stress that such contributions only arise because of our choice to formulate our description in terms of a bare basis, and we explicitly show in Sec. IV B that such terms `disappear' in the usual formulation in terms of a self-consistently dressed basis (where they become incorporated in the dressing of the bare trap eigenstates).

Unlike the evolution of total populations, the dynamics of condensed and thermal atoms will also  depend on processes of the general form $T_{ijji} T_{kllk}$ which will be intrinsically dispersive.  For simplicity, we shall henceforth limit our discussion to the case of only two partially condensed levels, namely $z_{0}, z_{1} \neq 0$, with level $2$ being the lowest purely incoherent level treated here.

\subsection{Evolution  of Condensate Populations}

The method used to obtain the evolution of condensate populations, is essentially the same as the one used for total populations (Eq. (\ref{FullPop2})).For ease of our subsequent comparison with the semi-classical laser theory, in this section we directly obtain equations for the modulus $|z_{i}|^{2}$ of the condensate mean field amplitude $z_{i}$ and not for $z_{i}$ itself. These are given by
\begin{equation}
\frac{d|z_{i}|^{2}}{dt} = - \sum_{rms}^{'} \left\{ \left[i  T_{ismr} z_{i}^{*}  \langle \hat{a}_{s}^{\dag} \hat{a}_{m} \hat{a}_{r} \rangle \right] + c.c. \right\} \label{z2}
\end{equation}
where the quantity analogous to $F_{i}$ of Eq. (5) is obtained from the equation of motion of the {\em entire} quantity $z_{i}^{*}  \langle \hat{a}_{s}^{\dag} \hat{a}_{m} \hat{a}_{r} \rangle$ and not {\em just} $\langle \hat{a}_{s}^{\dag} \hat{a}_{m} \hat{a}_{r} \rangle $ (see Appendix B). Thus, to second order in the effective interaction, we obtain 
\begin{equation}
\frac{d|z_{i}|^{2}}{dt} = -i \sum_{rms}^{'} \left[ T_{ismr} z_{i}^{*} \langle \hat{a}_{s}^{\dag} \hat{a}_{m} \hat{a}_{r} \rangle  -c.c. \right] + \left[ X_{ii}(T^{2}) + c.c. \right]
\end{equation}
with all $T^{2}$ collisional contributions summarized in $X_{ii}(T^{2})$, which can be written as 
\begin{equation}
X_{ii}(T^{2}) = R_{i}^{con} + S_{i} + E_{i} + \tilde{X}_{ii}(T^{2})
\end{equation}
$R_{i}^{con}$ corresponds to redistribution collisions affecting the level $i$ condensate, $S_{i}$ corresponds to self-interactions within level $i$, and $E_{i}$ gives the exchange collisions of level $i$ with its nearby levels, whereas the term $\tilde{X}_{ii}(T^{2}) $ corresponds to all off-diagonal contributions. For two condensed levels, in our simplified three-level system, the respective expressions are
\begin{eqnarray}
R_{0}^{con} = \Gamma_{0211} \left\{ \begin{array}{l}
2 |z_{0}|^{2} \left[ (n_{2}+1)n_{1}^{2} - n_{2}(n_{1}+1)^{2} \right] \\
+ 4 |z_{0}|^{2}|z_{1}|^{2} \left[ (n_{2}+1)n_{1} - n_{2}(n_{1}+1)     \right] \\
+|z_{0}|^{2}|z_{1}|^{4} \left[(n_{2}+1)-n_{2} \right]
 \end{array} \right\} 
\end{eqnarray} 
\begin{eqnarray}
S_{0} = - \lambda_{0000} \left[ \begin{array}{l}  2 |z_{0}|^{2} \left[  (n_{0}+1)^{2} n_{0} - n_{0}^{2} (n_{0}+1) \right] \\
+  |z_{0}|^{4} \left[ (n_{0}+1)^{2}-n_{0}^{2} \right] \end{array} \right]
\end{eqnarray}
\begin{eqnarray}
E_{0} & = & - 4 \lambda_{0110} \left[ \begin{array}{l}   |z_{0}|^{2} \left[ (n_{0}+1)(n_{1}+1)n_{1} - n_{0}n_{1} (n_{1}+1) \right] \\
+   |z_{0}|^{2} |z_{1}|^{2}  \left\{ \begin{array}{l} \left[ (n_{0}+1)(n_{1}+1) -n_{0}n_{1} \right] \\
+ \left[ (n_{0}+1)n_{1} -n_{0}(n_{1}+1) \right] \end{array} \right\}  \end{array} \right] \nonumber \\
& & -4 \lambda_{0220} |z_{0}|^{2} \left[ (n_{0}+1)(n_{2}+1)n_{2} - n_{0}n_{2} (n_{2}+1) \right]
\end{eqnarray}
and
\begin{eqnarray}
R_{1}^{con} = 
2 \Gamma_{0211} \left\{ \begin{array}{l} 
2  |z_{1}|^{2} \left[(n_{1}+1) n_{0}n_{2} - n_{1} (n_{0}+1)(n_{2}+1) \right] \\
+ 2 |z_{0}|^{2} |z_{1}|^{2} \left[ (n_{1}+1)n_{2} - n_{1}(n_{2}+1)     \right] \\
-|z_{1}|^{4} \left[ (n_{0}+1)(n_{2}+1) -n_{0}n_{2} \right] \\
-|z_{0}|^{2} |z_{1}|^{4} \left[ (n_{2}+1) - n_{2} \right]  \end{array} \right\} 
\end{eqnarray}
\begin{eqnarray}
S_{1} =  - \lambda_{1111} \left[ \begin{array}{l}  2 |z_{1}|^{2} \left[  (n_{1}+1)^{2} n_{1} - n_{1}^{2} (n_{1}+1) \right] \\
+  |z_{1}|^{4} \left[ (n_{1}+1)^{2}-n_{1}^{2} \right] \end{array} \right]
\end{eqnarray}
\begin{eqnarray}
E_{1} & = & - 4 \lambda_{0110} \left[ \begin{array}{l}   |z_{1}|^{2} \left[ (n_{0}+1)(n_{1}+1)n_{0} - n_{0}n_{1} (n_{0}+1) \right] \\
+    |z_{0}|^{2} |z_{1}|^{2}  \left\{ \begin{array}{l} \left[ (n_{0}+1)(n_{1}+1) -n_{0}n_{1} \right] \\
+ \left[ (n_{1}+1)n_{0} -n_{1}(n_{0}+1) \right] \end{array} \right\}  \end{array} \right] \nonumber \\
& & -4 \lambda_{1221} |z_{1}|^{2} \left[ (n_{1}+1)(n_{2}+1)n_{2} - n_{1}n_{2} (n_{2}+1) \right]
\end{eqnarray}
where
\begin{equation}
\lambda_{ijji} =  |T_{ijji}|^{2} \lim_{\eta \rightarrow 0^{+}}  \int_{0}^{\infty} d \tau e^{- \eta \tau }  
\end{equation}
while the expressions for the off-diagonal contributions $ \tilde{X}_{ii}(T^{2}) $ can be found in Appendix A.

\subsection{Evolution of Non-condensate Populations}

The evolution of the normal averages of levels 0 and 1 is governed by the equation
\begin{equation}
\frac{d}{dt}\langle \hat{c}_{i}^{\dag} \hat{c}_{i} \rangle  = - \sum_{rms}^{'} \left\{i T_{ismr} \left[ \langle \hat{a}_{i}^{\dag} \hat{a}_{s}^{\dag} \hat{a}_{m} \hat{a}_{r} \rangle - z_{i}^{*} \langle \hat{a}_{s}^{\dag} \hat{a}_{m} \hat{a}_{r} \rangle \right] -c.c. \right\} + \left[ Y_{ii}(T^{2}) + c.c. \right] \label{cc}
\end{equation}
where $Y_{ii}(T^{2}) = R_{i}^{th} -S_{i} -E_{i} + \tilde{Y}_{ii}(T^{2})$ and  $R_{i}^{th}$ are defined by 
\begin{eqnarray}
R_{0}^{th} = \Gamma_{0211} \left\{ \begin{array}{l} 
2 \left[ (n_{0}+1)(n_{2}+1)n_{1}^{2} - n_{0}n_{2}(n_{1}+1)^{2} \right] \\
+ 4 |z_{1}|^{2} \left[ (n_{0}+1)(n_{2}+1)n_{1} - n_{0}n_{2}(n_{1}+1) \right] \\ 
+ |z_{1}|^{4} \left[ (n_{0}+1)(n_{2}+1) - n_{0}n_{2} \right]
\end{array} \right\}
\end{eqnarray}
\begin{eqnarray}
R_{1}^{th} =  4 \Gamma_{0211} \left\{ \begin{array}{l} 
\left[(n_{1}+1)^{2} n_{0}n_{2} - n_{1}^{2} (n_{0}+1)(n_{2}+1) \right] \\
+  |z_{0}|^{2} \left[  (n_{1}+1)^{2} n_{2} - n_{1}^{2} (n_{2}+1) \right] \\  
+  |z_{1}|^{2} \left[(n_{1}+1) n_{0} n_{2}  - n_{1} (n_{0}+1)(n_{2}+1) \right]  \\  
+ |z_{0}|^{2}|z_{1}|^{2} \left[  (n_{1}+1) n_{2} - n_{1} (n_{2}+1) \right]
\end{array} \right\}
\end{eqnarray}
The evolution of the non-condensate component of level 2 is given by Eqs. (12)-(14), in the limit $z_{2}=0$. The respective expressions for the off-diagonal contributions $ \tilde{Y}_{ii}(T^{2}) $ can be found in Appendix A.

\section{Interpretation of Rate Equations}

We now turn our attention to the interpretation of the above equations, focusing in Sec. A only on the condensate evolution.  Ignoring, at first, re-distributional dynamics and off-diagonal contributions, we show how they reduce to the Hartree-Fock coupled nonlinear Schr\"{o}dinger equations conventionally used for describing two-component condensates (Sec. IV A1). By additionally considering the redistribution dynamics, we further discuss the analogy of a single inherently multi-mode condensate to the semi-classical photon laser theory \cite{Mandel}, with our discussion being limited to only two modes for clarity (Section IV A2). By further considering the evolution of uncondensed, and thus also of total populations of each level, we compare and contrast our approach to the kinetic theory of Walser et al. \cite{Walser_QK}, which is the formal multi-mode kinetic theory closest to the formalism of this paper (Sec. IV B).

\subsection{Coupled Condensate Evolution}

\subsubsection{Hartree-Fock Theory for Binary Condensates}

Let us initially forget about the presence of level 2 in our system and focus on the interactions between the two lowest levels in the trap (i.e. set $n_{2}=z_{2}=0$ and $V_{0211}, \lambda_{0220}, \lambda_{1221} \rightarrow 0$), which amounts to discussing two coupled partially-condensed systems.

Using Eqs. (17)-(18) and (20)-(21) we can deduce the evolution of the ground state condensate mean field $z_{0}$ (as opposed to $|z_{0}|^{2}$), which to second order reads
\begin{eqnarray}
\frac{dz_{0}}{dt} = -i \omega_{0} z_{0} & & -i \left[ T_{0000} -i \lambda_{0000} (1+2n_{0}) \right] \left( 2n_{0} + |z_{0}|^{2} \right) z_{0} + \lambda_{0000} \left( 2n_{0}^{2} \right) z_{0} \nonumber \\
& & -2i \left[ T_{0110} -2i \lambda_{0110} (1+n_{0}+n_{1}) \right] \left( n_{1} + |z_{1}|^{2} \right) z_{0} +2 \lambda_{0110} \left(2  n_{0}n_{1} \right) z_{0} \nonumber \\
& & -4 \lambda_{0110} \left( n_{1} - n_{0} \right) |z_{1}|^{2} z_{0} 
\end{eqnarray}
The terms in square brackets can be identified as the corresponding matrix elements of an effective {\em many-body} interaction introduced, strictly speaking, only over high-lying levels; this generalized effective interaction can be straightforwardly replaced by the usual many-body T-matrix $T_{ijji}^{MB}$ (over all levels), discussed in \cite{Stoof_NIST}, by subtracting from the second order expression of Eq. (29) (i.e. terms $\sim \lambda_{ijji}$), a term corresponding to the scattering of particles in vacuum, so that one ends up correctly calculating the {\em change} in the effective interaction when the pair of atoms collides in a condensed gas (as opposed to the vacuum) \cite{Morgan,Keith,Prouk_1D,Prouk_JPhysB}. Since our analysis has been carried out in terms of {\em bare} particle energies $\omega_{i}$, such a renormalization of the terms appearing in Eq. (29) merely amounts to neglecting the factor of 1 in $(1+2n_{0})$ and $(1+n_{0}+n_{1})$ \cite{Morgan_Rusch_Private}. We note that the same procedure should be carried out in all second order terms whose scattering amplitude depends on factors of the form $|T_{ijkl}|^{2} |z_{i}|^{2} |z_{j}|^{2} \left[ (n_{k}+1)(n_{l}+1) - n_{k} n_{l} \right]$ (if these are to be written in terms of $T^{MB}$), and such terms appear in Eqs. (20)-(24) and (27). After `renormalization', the square bracket in the first line of Eq. (29) takes the form
\begin{equation}
 T_{0000} -i \lambda_{0000} (1+2n_{0}) = T_{0000}^{2B} +T_{0000}^{2B} \lim_{\eta \rightarrow 0^{+}}  \int_{0}^{\infty} \frac{d \tau}{i} e^{- \eta \tau } \left(2n_{0} \right)  T_{0000}^{2B} = \left[ T_{0000}^{MB} \right]_{T^{2}}
\end{equation}
where $ \left[ T_{0000}^{MB} \right]_{T^{2}} $ corresponds to the second order expression for the many-body T-matrix; a similar identification can be made for the square bracket of the second line of Eq. (29) (where a factor of two arises from the symmetric interchange of atoms  $0$ and $1$ corresponding to direct and exchange Hartree-Fock terms). 

If one further ignores the last term of Eq. (29) (which corresponds diagrammatically to ignoring bubble diagrams with respect to the bare many-body ladder diagrams), one obtains the lowest order expression of the general equation
\begin{eqnarray}
i \frac{dz_{0}}{dt} =  \omega_{0}z_{0} & + & T_{0000}^{MB} \left( 2n_{0} + |z_{0}|^{2} \right) z_{0} -2   T_{0000}^{MB}  \left[  \lim_{\eta \rightarrow 0^{+}}  \int_{0}^{\infty} \frac{d \tau}{i} e^{- \eta \tau }  \right]  \left(  n_{0}^{2} \right)  T_{0000}^{MB} z_{0} \nonumber \\
& + & 2  T_{0110}^{MB} \left(  n_{1} + |z_{1}|^{2} \right) z_{0} - 2  T_{0110}^{MB} \left[  \lim_{\eta \rightarrow 0^{+}}  \int_{0}^{\infty} \frac{d \tau}{i} e^{- \eta \tau } \right]  \left(  2   n_{0} n_{1} \right)  T_{0110}^{MB} z_{0} \label{zT}
\end{eqnarray}
with the last term in each line ensuring correct scattering factors for condensate feed collisions from thermal atoms. This expression agrees with the equation for condensate evolution derived by one of us elsewhere \cite{Prouk_T_Matrix}. Although this equation is valid when dealing with a single condensed trap level, the somewhat heuristic neglect  of $-4 \lambda_{0110} \left( n_{1} - n_{0} \right) |z_{1}|^{2} z_{0}$ mentioned above suggests that it must be interpreted with some caution when the condensate spans more than one bare trap eigenstates \cite{Comment}.

We now show explicitly how the above equation reduces to the coupled finite temperature nonlinear Schr\"{o}dinger equations used for two-component condensation \cite{HF_Theory}. For this we must first assume that the gas is sufficiently dilute, so that we can further ignore in Eq. (31) the effect of the surrounding medium on binary collisions; this amounts to replacing $T_{0ii0}^{MB}$ ($i=0$, $1$) by the full two-body T-matrix $T_{0ii0}^{2B}$, while simultaneously ignoring the `kinetic' (many-body) contributions corresponding to the last term of each line in Eq. (32). By further approximating $T^{2B}$  by the usual pseudopotential $U_{ij} \delta({\bf r}-{\bf r^{'}})$ valid in 3D, where $U_{ij}$ is parametrized in terms of the scattering length for the collision of an atom in level $i$ with an atom in level $j$ ($i$,$j=0,1$) and transforming to coordinate space, we obtain
\begin{equation}
i \frac{\partial \Phi_{0}}{\partial t} = H_{0}^{(0)} \Phi_{0} + U_{00} \left( |\Phi_{0}|^{2} + 2 \tilde{n}_{0} \right) \Phi_{0} + 2 U_{01} \left( |\Phi_{1}|^{2} + \tilde{n}_{1} \right) \Phi_{0}
\end{equation}
where $\Phi_{i}$ and $\tilde{n}_{i}$ denote the condensate mean field and the non-condensate density of component $i$, with $H_{i}^{(0)}$ being the corresponding bare trap hamiltonian (kinetic energy plus trapping potential). By analogy,
\begin{equation}
i \frac{\partial \Phi_{1}}{\partial t} = H_{1}^{(0)} \Phi_{1} + U_{11} \left( |\Phi_{1}|^{2} + 2 \tilde{n}_{1} \right) \Phi_{1} + 2 U_{01} \left( |\Phi_{0}|^{2} + \tilde{n}_{0} \right) \Phi_{1}
\end{equation}
Eqs. (32)-(33) correspond to the well-known finite temperature Hartree-Fock equations for two-component condensation \cite{HF_Theory}.

\subsubsection{Two-mode Condensation vs. Semi-classical Laser Theory}

In the language of quantum optics, a system in which the condensate mean field spans more than one single-particle eigenstate should be analogous to a photon laser in multi-mode operation.  Exploring the formal connection between the two, with their similarities and differences, was in fact part of the motivation for this work and the main reason for formulating our approach in terms of bare particle eigenenergies.  If such a connection were to be taken literally, one would expect the coupled condensate mean field equations to resemble those of the two-mode laser. Even though we have assumed that level 2 is not itself condensed (and may even be initially unoccupied even by thermal atoms), we should still also consider its presence here; this is because the existence of level 2 is an inherent property of the system that cannot be ignored, since it will affect the evolution of the two lowest condensed modes $z_{0}$, $z_{1}$ via collisional redistribution processes. Combining Eqs. (19)-(24) for the second order contributions to the evolution of condensate amplitudes in a bare single-particle basis, they are found to exhibit the general structure
\begin{equation}
\frac{d |z_{0}|^{2}}{dt} = \alpha_{0}^{RSEH}  |z_{0}|^{2} - \beta_{0}^{S}  |z_{0}|^{4} - \theta_{01}^{RE} |z_{0}|^{2} |z_{1}|^{2} + \xi_{01}^{R} |z_{0}|^{2} |z_{1}|^{4}
\end{equation}
\begin{equation}
\frac{d |z_{1}|^{2}}{dt} = \alpha_{1}^{RSEH}  |z_{1}|^{2} - \beta_{1}^{RS}  |z_{1}|^{4} - \theta_{10}^{RE} |z_{0}|^{2} |z_{1}|^{2} + \xi_{10}^{R} |z_{0}|^{2} |z_{1}|^{4}
\end{equation}

Focusing initially only on the first three contributions of each of the above coupled equations, we note that they have the same form as those of the two-mode photon laser intensity equations \cite{Mandel}, if we  identify $|z_{i}|^{2}$ with the mode intensity. By analogy, we thus refer to the coefficients appearing in Eqs. (34)-(35) as: the `net gain' coefficient $\alpha_{i}$ of each mode, the `self-saturation' $\beta_{i}$  of each mode and the `cross-saturation' coefficients $\theta_{ij}$. In contrast to the optical laser where such equations arise from the polarization of the active medium, the nonlinearity in the case of the atom laser is intrinsic, arising from atom-atom interactions; this is absent in optical lasers, since there are no photon-photon interactions affecting the coherent photon field. Hence, the above coefficients $\alpha_{i}$, $\beta_{i}$ and $\theta_{ij}$ depend on incoherent populations $n_{i}$ and collisional rates $\lambda_{ijji}$ and $\Gamma_{0211}$. The superscripts $R$, $S$, $E$, $H$ used to define the above coefficients stand for Redistribution, Self, Exchange and Higher-level-exchange terms. The above coefficients are respectively defined by
\begin{equation}
\alpha_{i}^{RSEH} = \alpha_{i}^{R} - \alpha_{i}^{S} - \alpha_{ij}^{E} - \alpha_{ik}^{H}
\end{equation}
where

\begin{eqnarray}
\left\{ \begin{array}{l} \alpha_{0}^{R} = 2 \Gamma_{0211} \left[ (n_{2}+1) n_{1}^{2} - n_{2} (n_{1} + 1 )^{2} \right] \\
\alpha_{1}^{R} = 2 \Gamma_{0211} \left[ n_{0} n_{2} (n_{1}+1) - (n_{0}+1) (n_{2} + 1 ) n_{1} \right] \\
\alpha_{i}^{S} = 2 \lambda_{iiii} \left[ (n_{i}+1)^{2} n_{i} - (n_{i} + 1 ) n_{i}^{2} \right] \\
\alpha_{ij}^{E} = 4 \lambda_{ijji} \left[ (n_{i}+1) (n_{j}+1) n_{j}  - (n_{j}+1) n_{i} n_{j} \right]  \\
\alpha_{ij}^{H} = 4 \lambda_{ikki} \left[ (n_{i}+1) (n_{k}+1) n_{k}  - (n_{k}+1) n_{i} n_{k} \right] \end{array} \right\}
\end{eqnarray}

\begin{equation}
\beta_{i}^{(R)S} = \left( \beta_{i}^{R} \right) + \beta_{i}^{S}
\end{equation}
where $\beta_{0}^{R}=0$ and
\begin{eqnarray}
\left\{ \begin{array}{l} \beta_{1}^{R} = 2 \Gamma_{0211} \left[ (n_{0}+1) (n_{2}+1) - n_{0} n_{2}  \right] \\
\beta_{i}^{S} = \lambda_{iiii} \left[ (n_{i}+1)^{2} - n_{i}^{2} \right] \end{array} \right\}
\end{eqnarray}

\begin{equation}
\theta_{ij}^{RE} = \theta_{ij}^{R} + \theta_{ij}^{E}
\end{equation}
\begin{eqnarray}
\left\{ \begin{array}{l} \theta_{01}^{R} = -4 \Gamma_{0211} \left[ n_{1} (n_{2}+1) - (n_{1}+1) n_{2} \right] = -\theta_{10}^{R} \\
\theta_{ij}^{E} = 4 \lambda_{ijji} \left\{ \left[ ( n_{i}+1) (n_{j}+1) -  n_{1} n_{j} \right] +  \left[ ( n_{i}+1) n_{j} - (n_{j}+1) n_{i} \right] \right\} \end{array} \right\}
\end{eqnarray}

\begin{equation}
\xi_{01}^{R} =  \Gamma_{0211} \left[ (n_{2}+1) - n_{2} \right] = - \frac{1}{2} \xi_{10}^{R}
\end{equation}

Knowledge of their detailed form, enables us to draw important conclusions regarding the signs of the {\em total} coefficients $\alpha_{i}$, $\beta_{i}$, $\theta_{ij}$ and $\xi_{ij}$ of Eqs. (34)-(35). In particular we find, just as in the photon laser, the coefficients $\beta_{i} > 0$ always, thus giving rise to self-saturation of the mode intensity $|z_{i}|^{2}$, whereas coefficients $\alpha_{i}$ and $\theta_{ij}$ can be positive or negative (depending on the values of $n_{i}$, $\lambda_{ijji}$ and $\Gamma_{0211}$). The net gain coefficient of each mode is given by Eq. (36); here $\alpha_{i}^{R}$ gives the gain coefficient due to redistribution collisions $|T_{0211}|^{2}$ and can be positive or negative, depending on the relative values of $n_{0}$, $n_{1}$ and $n_{2}$. The remaining contributions to Eq. (36) arise from collisions of an atom in level $i$ with an atom in level $j$ or $k$ (where $i=0,1$ and $j= 0,1,2$, while $k \neq i, j$), and leads to saturation of intensity growth since $\alpha_{i}^{S}, \alpha_{i}^{E},  \alpha_{i}^{H} > 0$ always. The cross-saturation rates $\theta_{ij}$ contain contributions from exchange and redistribution collisions and can be positive or negative.

An important difference between Eqs. (34)-(35) and the corresponding ones for two-mode photon lasers is the presence of the higher order cross-saturation terms $\xi_{ij} |z_{0}|^{2} |z_{1}|^{4}$, which arise solely as a result of collisional redistribution. In deriving the two-mode photon laser equations, one conventionally performs a perturbative expansion of the polarization of the active medium in terms of laser intensity. For near-threshold operation and weak laser intensities, the third order perturbative expansion of the medium polarization is usually sufficient, and such a truncation generates at most terms proportional to the square of the laser intensity. Terms of third order in the laser intensity (as well as higher order ones) would indeed arise in two-mode photon laser theory, if one extended the perturbative treatment  of the polarization to fifth order (or beyond).

In the case of Bose-Einstein condensation there is no active medium to be polarized, and the nonlinearity of the system is due to intrinsic atom-atom interactions. In this case, the highest order $|z_{0}|^{2m} |z_{1}|^{2n}$ to be included in the two-mode equations for the `intensities' $|z_{i}|^{2}$ is based on the number of single-particle operators appearing in the nonlinear interaction term in the hamiltonian of the system, i.e. whether one includes only two-body collisions (as usual, in the dilute limit $n a^{3} \ll 1$), or also three-body collisions (or higher). In the usual case of two-body collisions as defined by Eq. (1), the resulting equations for $|z_{i}|^{2}$ may contain terms up to order $(z^{*}z)^{3}$; so, in general, it would not appear justified to ignore contributions of third order in the mode intensities. We note that the coefficients of the higher order cross-saturation terms appearing in Eqs. (34)-(35) have opposite signs and are given by  $\xi_{10} = - 2 \xi_{01}$, where  $\xi_{01}=\Gamma_{0211} > 0$.

The above discussion has been given in terms of bare amplitudes $|z_{i}|^{2}$, and we should note that the form of these equations changes when the $|z_{i}|^{2}$ refer to amplitudes of dressed modes. In particular, when the $|z_{i}|^{2}$ refer to amplitudes in modes dressed only by the condensate mean field (assuming it is physically meaningful to speak of more than a single condensed mode in such a dressed basis), such higher order cross-coupling contributions would not arise, thus yielding a direct analogy with the intensity equations for two-mode photon lasers. The justification for this is given in the next section, where we discuss how the second order collisional integrals become modified, upon shifting our  single-particle eigenstates to a basis dressed by mean fields. The `exact' analogy between two-mode condensation and optical lasers arising in this case might suggest that even in a basis dressed by the condensate mean field, one should, in principle, deal with more than one condensed modes in a trapped assembly. This analogy of a single, inherently multi-mode, condensate with the usual semi-classical multi-mode 
photon laser theory, however, appears to break down when shifting to a basis including higher mean field effects (i.e. those due to uncondensed atoms, anomalous averages, etc.)

In this section, we have presented some similarities and differences between two-mode Bose-Einstein condensation  and semi-classical two-mode photon laser theory. A more detailed investigation should discuss such equations in the presence of pumping, evaporative cooling and coherent outcoupling (in a manner analogous to single-mode atom laser models \cite{AL1,AL2,AL3,AL4,Atom_Laser_Review}). 
The first obvious modification that would occur in this case is that the `net gain' coefficient(s) of the lasing mode(s) should become positive and large, since the contribution of an irreversible evaporative cooling mechanism combined with the redistribution collisions should result in a large flow of particles towards such mode(s). At the same time, of course, the $\alpha_{i}$ coefficients will acquire an additional negative contribution whose value will depend on the rate of outcoupling of atoms from the particular condensed level. This should enable the system to reach a steady state, with the coherent amplitude growth being stabilized both by the outcoupling mechanism, as well as by the collisionally-induced dephasing \cite{Horak} due to self ($\lambda_{iiii}$) and exchange  ($\lambda_{ijji}$) interactions. In our formalism, such dephasing can be viewed as destruction of the coherent mean field amplitude $|z_{i}|^{2}$ in favour of the `incoherent populations' $n_{i}$ (i.e. transfer of population from `coherent' to `incoherent'). Inherent two-body and three-body inelastic loss processes, which have recently been shown to be essential for reaching a steady state for an atom laser \cite{Robins}, will affect the coherent and incoherent populations of a particular mode in a different manner due to the nature of our mean field decorrelation, e.g. $\langle \hat{a}_{i}^{\dag}  \hat{a}_{i}^{\dag}  \hat{a}_{i}^{\dag} \hat{a}_{i} \hat{a}_{i}  \hat{a}_{i}  \rangle \sim \left( |z_{i}|^{6} + 6 n_{i}^{3} \right)$, thus creating  an essential irreversibility for the condensate mean field $|z_{i}|^{2}$ to dominate over its corresponding fluctuations $n_{i}$. By only outcoupling the lowest condensed mode of the system, one could, for example, achieve a kind of `population inversion', in the sense that the largest condensate particle number accumulation might occur in some state other than the ground state of the bare trap. In closing this subsection, we note that the quantum nature of the multi-mode atom laser is implicit in the general Fokker-Planck treatment developed by Stoof \cite{Stoof_FP}.

\subsection{Links to Other Kinetic Theories}

The expressions given earlier for the evolution of condensed and uncondensed components contain terms in the condensate mean field beyond order $(z^{*}z)$, and at first sight this appears to be in disagreement with existing kinetic theories \cite{Stoof_FP,Walser_QK,Griffin_Hydro,Milena,Stoof_Growth}. For example, comparing  our final expressions for total populations to those of Walser et al. \cite{Walser_QK} (in the corresponding limit of diagonal populations and no anomalous averages),  we appear to obtain an excess $\Delta N$ for the populations of levels 0 and 2 (whereas level 1 is underestimated by twice that amount), with  $\Delta N$ given by
\begin{equation}
\Delta N =  2\Gamma_{0211} |z_{1}|^{2} \left\{ \begin{array}{l}  
 4 |z_{0}|^{2}  \left[ (n_{2}+1) n_{1} - n_{2} (n_{1}+1) \right] \\ 
+ |z_{0}|^{2} |z_{1}|^{2} \left[ (n_{2}+1)- n_{2} \right] \\ 
+ |z_{1}|^{4} \left[ (n_{0}+1)(n_{2}+1)-n_{0}n_{2} \right] 
 \end{array} \right\}
\end{equation}
Since all terms of Eq. (43) are proportional to $|z_{1}|^{2}$, it is clear that  our equations will assume the usual form (e.g. as discussed in Walser et al. \cite{Walser_QK}) in the limit of extremely weak condensation (i.e. when only the lowest trap eigenstate is occupied by the condensate). The natural question arising then is whether such equivalence remains beyond this simple limit.

The key to understanding this apparent inconsistency is to note that our treatment has so far been given in terms of single-particle eigenfunctions, whose energies are bare (unshifted) ones, as if the trap were completely void. However, since the trap contains a large atomic medium which is condensed, the collisions will actully occur in the presence of the condensate, which forces the eigenenergies to  vary in time due to the existing mean field potentials. To account for this, one can thus convert  from the simple picture in terms of unshifted energies employed above, to the conventional one in which the effects of the mean fields are included into suitably `renormalized'  basis eigenenergies \cite{QKV,Stoof_FP,Walser_QK}. In the next section we show explicitly that shifting our basis in a suitable manner identically reproduces the kinetic theory of Walser et al. \cite{Walser_QK}, thus providing an alternative derivation of the latter theory.

To understand the relation of our theory to existing kinetic treatments, we must, in first instance, explain the physical origin of all terms of Eq. (43). We recall from Eq. (3) that all collisional contributions affecting the total population of a particular level $i$ arise from terms of general structure $\left[ \sum_{rms} T_{ismr} \langle \hat{a}_{i}^{\dag} \hat{a}_{s}^{\dag} \hat{a}_{m} \hat{a}_{r} \rangle + h.c. \right] $. Defining in the usual manner \cite{Prouk_NIST,Blaizot,Thesis} the quantities $\rho_{ji}=\r{i}{j}$ and $\kappa_{jk}=\k{k}{j}$ and focusing initially on the first line of Eq. (43), we note that  terms  $|z_{0}|^{2} |z_{1}|^{2}$ arise from the elimination of $\rho$ appearing in  $\langle a^{\dag} a^{\dag} a a \rangle \sim \left[ 4 \rho (z^{*} z) + c.c. \right] $, which, among other contributions leads to
\begin{equation}
\frac{d N_{i}}{dt} = - 4 \sum_{rms} T_{ismr} z_{i}^{*} \left\{ \sum_{klt}  T_{mltk}^{(\delta)} \left[ \rho_{ks} (z_{l}^{*}z_{t}) \right] - T_{klts}^{(\delta)} \left[ \rho_{mk} (z_{l}^{*}z_{t}) \right] \right\} z_{r} + c.c.
\end{equation}
Here the notation $T^{(\delta)}$ stands for the corresponding two-body t-matrix multiplied by the approximately energy-conserving integral of Eq. (10), and the above term clearly contains a contribution $\sim \left\{ |T_{0211}|^{2} z_{0}^{*} \left[ (n_{1}-n_{2}) (z_{1}^{*} z_{0}) \right] z_{1} + c.c. \right\}$ These latter  terms would obviously not arise if the quantity $\rho$ was considered essentially constant, as might be more appropriate, for example, for a rapidly-thermalizing `hydrodynamic' system.
Similarly, terms of order $|z_{i}|^{4}$ (i.e. second line of Eq. (43)) arise from the elimination of the pair anomalous average $\kappa$ in $ \langle a^{\dag} a^{\dag} a a \rangle \sim  \left[ \kappa (z^{*} z^{*}) + c.c. \right] $, yielding (among other terms)
\begin{equation}
\frac{d N_{i}}{dt} = - \sum_{rms} T_{ismr} \left( z_{i}^{*} z_{s}^{*} \right) \left\{ \sum_{pq} \left[ T_{mrpq}^{(\delta)} + \sum_{l} \left(  T_{mlpq}^{(\delta)} \rho_{rl} +  V_{lrpq}^{(\delta)} \rho_{ml} \right) \right] \left( z_{p} z_{q} \right) \right\} + c.c.
\end{equation}
which leads to $\sim \left\{  |T_{0211}|^{2} ( z_{1}^{*} z_{1}^{*}) (1+n_{0}+n_{2}) (z_{1} z_{1}) + c.c. \right\} $
This proves that our expressions explicitly contain anomalous condensate terms $(zz)$ and $(z^{*}z^{*})$, in contrast to the non-condensate anomalous average $\kappa$ which has been neglected from all final expressions due to the application of the Popov approximation.

Finally, terms of order  $(z^{*}z)^{3}$ (last contribution of Eq. (43)) can only arise from elimination of $z$ (or $z^{*}$) wherever it appears in the first order expressions, thus yielding
\begin{eqnarray}
\frac{d N_{i}}{dt} & & = - \sum_{rms} T_{ismr} z_{i}^{*} \left\{ z_{s}^{*} z_{m} \left[ \sum_{pql}  T_{rlpq}^{(\delta)} (z_{l}^{*}z_{p}z_{q}) \right] + z_{s}^{*} \left[  \sum_{pql}  T_{mlpq}^{(\delta)} (z_{l}^{*}z_{p}z_{q}) \right] z_{r} \right\} \nonumber \\
& & + \sum_{rms} T_{ismr} \left\{ z_{i}^{*} \left[ \sum_{pql}  T_{pqls}^{(\delta)} (z_{p}^{*}z_{q}^{*}z_{l}) \right] z_{m} z_{r}  + \left[ \sum_{pql}  T_{pqli}^{(\delta)} (z_{p}^{*}z_{q}^{*}z_{l}) \right] (z_{s}^{*}  z_{m} z_{r}) \right\} + c.c.
\end{eqnarray}
where the products $(z^{*}z^{(*)}z)$ appearing immediately after each $T_{\cdots}^{(\delta)}$ are now the ones arising from the elimination of a $z$ or $z^{*}$ from the first order expression in $T (z^{*}z^{*}zz)$. This allows us to interpret the term $\sim |T_{0211}|^{2} |z_{0}|^{2} |z_{1}|^{4}$ of Eq. (43) as arising from the terms  $(z^{*}zz)$ generated in second order expressions, and these could be loosely interpreted as arising either (i) from condensate anomalous averages of the form $(z_{i} z_{j})$, or (ii) from off-diagonal coherent terms $(z_{i}^{*} z_{j})$, with the two statements being equivalent and showing that it would not be justified to include off-diagonal  $(z_{i}^{*} z_{j})$, while ignoring condensate anomalous averages $(z_{i} z_{j})$.

In short, the above analysis shows clearly that all terms  $(z^{*}z)$, $(zz)$ and $(z^{*}z^{*})$ of second or higher order appearing in Eqs. (13), (17), (26) and hence (43) can only be generated by the adiabatic elimination of quanitites $z$, $\rho$ and $\kappa$ from the corresponding first order expressions. In particular, the terms of Eq. (43) respectively arise from the adiabatic elimination of the quantities $\rho_{12}$, $z_{2}$ and $\kappa_{02}$. Such elimination is fully justified in a bare basis, which assumes that all quantities evolve in an analogous (rapid) fashion and should thus be treated consistently. However, tranformation to a dressed basis automatically restricts certain quantities to be slowly-evolving (which parameters are slowly-evolving depends explicitly on the choice of unperturbed basis dressing the single-particle eigenstates), so that their elimination can no longer be justified. In the next section we thus extend the above discussion to show explicitly that our treatment yields precisely the second order collisional integrals of Walser et al. \cite{Walser_QK}, upon renormalizing our single-particle eigenenergies to those dressed by HFB mean fields.

\subsubsection{Equivalence to Theory of Walser et al.}

To show explicitly the link of our treatment to the kinetic theory of Walser et al. \cite{Walser_QK}, we now focus on the general expression for the normal uncondensed component in an n-level system. To establish exact analogy, in this section we further incorporate the uncondensed anomalous average $\kappa$ into our treatment, since this is explicitly present in the expressions of Walser et al. \cite{Walser_QK}.

After much algebraic manipulation of the second order contributions to Eq. (26), we hence obtain the following result for the second order collisional integrals arising within our bare basis
\begin{eqnarray}
\left[ \frac{d}{dt}(\rho_{ii}) \right]_{T^{2}}^{Bare} = \left[ \frac{d }{dt} (\rho_{ii}) \right]_{T^{2}}^{Walser} + \left( \sum_{rms} T_{ismr}\left\{ \begin{array}{l} \left( 2 \rho_{ms} \rho_{ri}  + \kappa_{mr} \kappa_{is}^{*} \right) \\ + \left(2  \rho_{mi} z_{s}^{*} z_{r} + z_{m} z_{r} \kappa_{is}^{*} \right) \end{array} \right\} + h.c. \right)
\end{eqnarray}
The first term in the above expression has exactly the same form as the second order collisional integrals obtained  by Walser et al. \cite{Walser_QK} (with the only difference being that in our approach the eigenenergies of Eq. (10) are now explicitly given by their bare values $\omega_{i}$), while the second term yields all the additional terms of our treatment arising when working within a bare-particle basis. The terms within the curly bracket appearing in the latter term denote the corresponding second order contributions arising from the adiabatic elimination of the quantities within the brackets, in the particular basis chosen. Their explicit form is obtained by the formal solution of their respective equations of motion given below, and such second order contributions therefore implicitly include the energy-conserving integral of Eq. (10). In the full hamiltonian of Eq. (1), the equations of motion for the above quantities are given by \cite{Prouk_NIST,Thesis}
\begin{equation}
i \hbar \frac{dz_{n}}{dt}  = \sum_{k} \Xi_{nk}^{Bare} z_{k} + \sum_{ijk} T_{nijk} \left[ z_{i}^{*}z_{j}z_{k} + \kappa_{jk}z_{i}^{*} + 2\rho_{ji}z_{k} \right] \label{z} 
\end{equation}
\begin{eqnarray}
i \hbar \frac{d \rho_{ji} }{dt} & = & \sum_{n} \left( \Xi_{jn}^{Bare} \rho_{ni} - \Xi_{ni}^{Bare}  \rho_{jn} \right)  + \sum_{r} \left[ \eta_{jr} \rho_{ri} - \rho_{jr} \eta_{ri} \right] \nonumber \\
&  & - \sum_{r} \left[\kappa_{jr}  \Delta_{ri}^{*}  - \Delta_{jr}
\kappa_{ri}^{*} \right]
\end{eqnarray}
\begin{eqnarray}
i \hbar \frac{d \kappa_{kj}}{dt} & = & \sum_{n} \left( \Xi_{jn}^{Bare} \kappa_{nk}  + \Xi_{kn} \kappa_{jn} \right)+ \sum_{s} \left[ \eta_{ks} \kappa_{sj}  + \kappa_{ks}  \eta_{sj}^{*}
\right] \nonumber \\
& & + \Delta_{kj} + \sum_{s} \left[ \rho_{ks} \Delta_{sj}  + \Delta_{ks}
\rho_{sj}^{*}  \right]  
\end{eqnarray}
where we have ignored in the final expressions higher order correlations such as $\langle \hat{c}^{\dag} \hat{c} \hat{c} \rangle$ and  $\langle \hat{c} \hat{c} \hat{c} \rangle$ which arise in the above expressions (their effect will be discussed elsewhere). In the above expressions, we have further defined $\eta_{pq} =  2 \sum_{kl} T_{pklq} \left[z_{k}^{*} z_{l}  + \rho_{lk} \right] $
and $\Delta_{pq} =  \sum_{kl} T_{pqkl}   \left[ z_{k} z_{l} + \kappa_{kl}  \right] $.
Choosing to describe our system in terms of bare trap eigenenergies essentially amounts to diagonalizing the bare trap hamiltonian $H_{0} = \sum_{ij} \Xi_{ij}^{Bare} \langle \hat{a}_{i}^{\dag} \hat{a}_{j} \rangle$. In that case, the energies $\omega_{i}$ appearing in the energy-conserving condition (10) correspond to the eigenstates of the harmonic trap.

However, instead of working with a bare basis, one can choose a dressed `unperturbed' basis $(H_{0}+H_{Q})$ in which to define the renormalized single-particle eigenenergies, via
\begin{equation}
\hat{H} = \left( H_{0}+ H_{Q} \right) + \left( V - H_{Q} \right)
\end{equation}
where $V$ corresponds to the binary collision hamiltonian defined in Eq. (1) and $H_{Q}$ denotes the quasiparticle hamiltonian
\begin{equation}
H_{Q} = \frac{1}{2}  \sum_{pq} \left\{ h_{pq} \left( \hat{c}_{p}^{\dag}
\hat{c}_{q} + \hat{c}_{q} \hat{c}_{p}^{\dag} \right) + \left( \Delta_{pq}
\hat{c}_{p}^{\dag} \hat{c}_{q}^{\dag} + \Delta_{pq}^{*} \hat{c}_{q} \hat{c}_{p}
\right) \right\} \label{QHam}  
\end{equation}
with $h_{qp} = \sum_{pq} \left( \Xi_{pq}^{Bare} + \eta_{pq} \right) $. In this case, one is  incorporating part of the collisional evolution of Eqs. (48)-(50)  into their `free evolution frequencies' $\omega_{i}$, which now become dressed to $\tilde{\omega}_{i}$. We thus find that when working in the usual quasiparticle basis, and upon ignoring higher order (triplet) averages, the above equations for $z$, $\rho$ and $\kappa$ simplify to $(dz_{n}/dt) = -i \tilde{\omega}_{n} z_{n}$, $ (d \rho_{ji}/dt) = -i \left( \tilde{\omega}_{j} - \tilde{\omega}_{i} \right) \rho_{ji} $ and $( d\kappa_{jk}/dt) = -i \left( \tilde{\omega}_{j} + \tilde{\omega}_{k} \right) \kappa_{jk}$
where  the $\tilde{\omega}_{i}$ now correspond to the renormalized eigenenergies in the above chosen basis. Hence (assuming as usual that initial interparticle correlations can be ignored), the quantities $z$, $\rho$ and $\kappa$ lead to no intermediate collisional evolution (to lowest order), which  shows clearly that all terms of second or higher order in $(z^{*}z)$ or $(zz)$ indeed vanish when dealing with single-particle eigenenergies dressed by the HFB mean fields; in the latter case our theory conincides with that of Walser et al (who actually formulated their theory in terms of unspecified number-conserving renormalized potentials as can be seen from Eqs. (64)-(67) of \cite{Walser_QK} ).

Furthermore, this treatment brings out the implicit basis dependence of the second order collisional integrals of Walser et al. In their treatment, they assume that the evolution of the system can be well parametrized by a restricted set of slowly-varying `master' variables. As such, the possibility of evolution of such variables is absent from their theory, and hence they obtain a basis-independent formulation. Our treatment, however, indicates that their choice of a renormalized single-particle basis is implicit in their particular  choice of the slowly-varying quantities. In a single-particle basis where the eigenenergies are dressed by HFB mean fields $z$, $\rho$ and $\kappa$, the collisional integrals cannot include contributions due to the variations of these HFB parameters which are assumed to be static and hence our treatment identically coincides with that of Walser et al., defining at the same time the eigenenergies $\tilde{\omega}_{i}$ as being dressed by HFB self-consistent potentials. This should be contrasted to the dressing caused simply by number-conserving HF mean fields which is implicit in Eq. (50) of their treatment. However, if one works within a bare basis, there is no a priori reason to assume that any quantity is slowly varying with respect to the others, and this leads to the additional second order contributions we have obtained in this paper (while simultaneously restricting the eigenstates of Eq. (10) to those of the harmonic trap). Hence, we believe our analysis further confirms the statement made in the last sentence of their paper, that their results should be valid when the mean-field induced energy shifts may be neglected during a strong collisional event, and we thus interpret the basis correction terms of our treatment as the evolution of these shifts in a {\em bare} basis.  Our treatment can be readily generalized to all basis-dependent shifts, by respectively defining other dressed basis as follows: (i) the $T=0$ Gross-Pitaevskii basis in which only the condensate parameters $z$ evolve slowly and (ii) the Hartree-Fock basis in which both $z$ and $\rho$ are taken as slowly-varying, whereas the anomalous average $\kappa$ is adiabatically eliminated (even over low-lying\footnote{We remind the reader that the corresponding values of all quantities over high-lying modes have already been implicitly adiabatically eliminated, in favour of the effective two-body interaction $T$ over low-lying modes used throughout this work.} modes). Finally, in the case of an HFB formulation, such extra terms will only arise from triplet and higher-order averages ignored in Eqs. (48)-(50) and will clearly be of higher (than second) order in the interatomic potential, as also implied by Walser et al. \cite{Walser_QK}.

\section{Conclusions}

In this paper we have discussed the dynamics of the lowest-lying single-particle bare trap levels in the weak coupling limit, under the assumption that more than one level exhibit condensation.  By comparing our equations to those of the two-mode photon laser, we briefly commented on the similarities and differences of such equations. We note that the usual discussion of two-mode condensation focuses on  condensates which are either in different spin states, or under physical separation. Describing each of the two separate condensates by a nonlinear Schr\"{o}dinger equation (as discussed in Sec. IV), one indeed recovers a direct analogy to the semi-classical theory of two-mode photon lasers (by associating the fields $|\Phi_{i}(r,t)|^{2}$ to the laser intensity). However, since the multi-mode nature of the photon laser refers to different {\em modes} of the electromagnetic field, we believe that a more direct analogy can be obtained at a more fundamental level, namely within a {\em single} trapped condensate (see also \cite{Stoof_FP}). In particular, in the case of photon lasers, the number of modes and the precise equations governing them depend on the atomic species, the cavity, the intensity and the truncation of the polarization expansion imposed. In the end, the modes into which the field will oscillate result from the strong coupling of the active medium to the cavity, corresponding thus, in some sense, to a dressed basis. 
In the case of a trapped Bose-Einstein condensate, the number of modes spanned by the condensate is determined by its size, its diluteness (i.e. whether the discussion can be essentially limited to two-body interactions at the particular temperatures and densities), the strength of the interactions and the trap confinement power, whereas the form of the equations additionally depends on the basis in which we choose to describe the system. In this case, the maximum factor $|z_{i}|^{2m} |z_{j}|^{2n}$ which can be obtained in the equations is set by the complexity of the many-body interactions (i.e. two-body, three-body, etc.), whereas the factor we actually find in our final equations depends on the basis in which the equations are explicitly formulated. Following our renormalization discussion of Sec. V B, we stress that in the case of binary interactions, exact analogy with the semi-classical two-mode photon laser theory is obtained {\em only} within a $T=0$ Gross-Pitaevskii basis, i.e. a basis dressed only by the condensate mean field potentials via $H_{0}^{'} = \sum_{rn} \Xi_{rn}^{Bare} \langle \hat{a}_{r}^{\dag} \hat{a}_{n} \rangle + (1/2) \sum_{rsmn} T_{rsmn} z_{r}^{*} z_{s}^{*} z_{m} z_{n} $, in analogy to the above-mentioned behavior of the photon laser.

Changing our description slightly to explicitly include the mean-field effects on eigenenergies of our single-particle system, we demonstrated that our treatment is consistent with the usual description in which condensed atoms are described in terms of the NLSE, and the evolution of non-condensed atoms is based on the quantum Boltzmann equation \cite{Walser_QK,Griffin_Hydro,Stoof_Growth}. In this case, the single-particle eigenenergies are ultimately effectively dressed by mean fields in the many-body T-matrix approximation, just as was found in the approach of Stoof \cite{Stoof_FP} (with the many-body corrections arising from suitable inclusion of the anomalous average \cite{Prouk_T_Matrix,Morgan,Prouk_1D,Prouk_JPhysB}). Our approach has been compared in more detail to the treatment of Walser et al \cite{Walser_QK}, which has been recently shown \cite{Wachter} to be equivalent to the Kadanoff-Baym Green's function formalism as applied to trapped Bose gases by Imamovic-Tomasovic and Griffin \cite{Milena}. In particular, we have discussed how additional contributions which arise in our second order collisional integrals upon choosing a simpler (than HFB) unperturbed basis, depend on the choice of this (bare or partially dressed) basis. This suggests an implicit assumption by Walser et al. that their eigenenergies are dressed by mean fields in the Hartree-Fock-Bogoliubov approximation, which is equivalent to their choice of slowly-evolving master variables. Our method thus yields an alternative microscopic derivation of the theory of Walser et al., based on a coupled equations of motion formalism.

When applying perturbation theory only to second order in the potential, one has to justify why such a truncation is physically realistic. This perturbation theory is essentially a systematic expansion in terms of the diluteness parameter $\sqrt{n a^{3}}$. At $T=0$ such a treatment can be justified in the limit $\sqrt{n a^{3}} \ll 1$ \cite{Diluteness}. For a large system leading to large mean fields which heavily dress the single-particle eigenenergies from their bare trap values, one conventionally shifts to a description in terms of quasiparticles, thus employing HFB-shifted eigenenergies. In this case, the validity criterion of perturbation theory at finite temperature essentially becomes $ (kT/nU_{o})(\sqrt{na^{3}}) \ll1$ for the homogeneous system \cite{PT,Gora,Morgan}, a criterion closely related to the one for the absence of critical fluctuations occuring sufficiently close to the transition point\cite{Ginzburg}. On the other hand, the main part of this paper has been based on a treatment in terms of  bare trap eigenenergies, since this enables a simple and direct analogy with the semi-classical equations for multi-mode photon laser theory. Clearly such a treatment cannot be valid for large, dense condensates, and its validity will be restricted to systems in which the interactions and particle numbers are so small, that the trap eigenenergies become only slighlty perturbed by the mean fields. A minimum (but not necessarily sufficient) criterion here is that the condensates are very weakly-interacting in the sense that $nU_{o} \ll \hbar \omega$. By identifying the additional basis-dependent corrections of our treatment to the second order collisional integrals of Walser et al, we can restate this condition as the requirement that the additional terms are much smaller than the first order contributions. Since the additional terms can be visualized as rates of change of the parameters $z$, $\rho$ and $\kappa$, this criterion essentially reduces to the slow evolution of such quantities. Hence, our bare basis analysis can only be useful in the limit when all mean field potentials evolve very slowly, and do not heavily modify the bare trap eigenenergies. However, our explicit expressions additionally include such slow evolution from a {\em bare basis} description, as opposed to existing theories in which such evolution is absent, applicable in the domain where the mean field energy shifts induced (on the already renormalized eigenenergies) during a collision can be neglected. Indeed, by re-formulating our treatment explicitly in terms of HFB eigenenergies, we find contributions of this type arising only in higher orders in the potential, being generated by the careful consideration of triplet and higher order averages, an issue which will be explicitly addressed elsewhere.

\acknowledgements
We acknowledge discussions with R. Walser and J. Williams. This work was partially supported by the ULF (Contract No. ERB-FMGECT 950021) and by a grant of the Max-Planck-Institute for Quantum Optics, Garching, where NPP originally developed part of this formalism.

\appendix \section{Off-Diagonal Contributions}

For completeness, we give here all off-diagonal contributions to the condensate / non-condensate populations of the three-level system discussed in the text. In particular, we have for the total population evolution

\begin{eqnarray}
\tilde{R}_{T_{|0211|^{2}}} = & &  4 \Gamma_{0211}  \left\{ \begin{array}{l}  (N_{1}-N_{2}) \rho_{01} \rho_{10} + (N_{1}^{T}-N_{2}) (\zeta_{01} \rho_{10} + \rho_{01} \zeta_{10}) \\ +  (N_{1}-N_{0}) \rho_{12} \rho_{21} + (N_{1}^{T}-N_{0}) (\zeta_{12} \rho_{21} + \rho_{12} \zeta_{21} ) \end{array} \right\} \nonumber \\
& & -2 \Gamma_{0211}  (1+2 N_{1}) \left[ \rho_{02}\rho_{20} + \zeta_{02}\rho_{20} + \rho_{02} \zeta_{20} \right]  \nonumber \\
& & -8 \Gamma_{0211} \left\{ \rho_{01} \rho_{12} \rho_{20} +   \zeta_{01} \rho_{12} \rho_{20} +  \rho_{01} \zeta_{12} \rho_{20} + \rho_{01} \rho_{12} \zeta_{20} \right\} \label{R}
\end{eqnarray}

\begin{eqnarray}
\tilde{Q}_{ijji}^{0211} = & &  - \nu_{01}^{(2)}  \left\{ \begin{array}{l}  \rho_{10} \left[  4  N_{0} \rho_{12} +  4 N_{0}^{T}\zeta_{12} +2\left( \rho_{10} \rho_{02}+ \zeta_{10} \rho_{02}+ \rho_{10} \zeta_{02} \right) \right] \\ + \zeta_{10} \left[  4 N_{0}^{T}\rho_{12} + 2 N_{0}^{T}\zeta_{12} +\left( 2 \rho_{10} \rho_{02}+ \zeta_{10} \rho_{02}+ 2 \rho_{10} \zeta_{02} \right) \right] \end{array} \right\} \nonumber \\
& & - \nu_{21}^{(0)}  \left\{ \begin{array}{l}  \rho_{12} \left[  4  N_{2} \rho_{10} +  4 N_{2}^{T}\zeta_{10} +2\left( \rho_{12} \rho_{20}+ \zeta_{12} \rho_{20}+ \rho_{12} \zeta_{20} \right) \right] \\ + \zeta_{12} \left[  4 N_{2}^{T}\rho_{10} + 2 N_{2}^{T}\zeta_{10} +\left( 2 \rho_{12} \rho_{20}+ \zeta_{12} \rho_{20}+ 2 \rho_{12} \zeta_{20} \right) \right] \end{array} \right\} \nonumber \\
& & +6 \vartheta_{01}^{(2)} \left[ 2 N_{1} \rho_{10} \rho_{12} + 2N_{1}^{T} \left( \zeta_{10} \rho_{12} + \rho_{10} \zeta_{12} \right)+ N_{1}^{TT} \zeta_{10} \zeta_{12} \right] \nonumber \\
& & +  \varepsilon_{01}^{(2)} \left[ 2 \rho_{10} \rho_{12} + 2 \left( \zeta_{10} \rho_{12} + \rho_{10} \zeta_{12} \right)+ \zeta_{10} \zeta_{12} \right]
\end{eqnarray}
Although $\tilde{R}_{T_{|0211|^{2}}}$ vanishes in the limit of diagonal $\rho$, the same does not apply to $ \tilde{Q}_{ijji}^{0211} $ which  contains a contribution $\sim \zeta_{10} \zeta_{12}$, which will only vanish upon assuming that one of the three levels (e.g. level 2) is fully uncondensed. 

In the above expressions, in addition to the total population $N_{i}=n_{i}+|z_{i}|^{2}$, we have defined the `population terms' 
$N_{i}^{T}  =  \frac{1}{2} \left( 2n_{i} + |z_{i}|^{2} \right)$ and
$N_{i}^{TT}  =  \frac{1}{3} \left( 3n_{i} + |z_{i}|^{2} \right) $.
The population $N_{i}^{T}$ appears familiar from laser physics, as it contains the thermal (chaotic) contribution with a pre-factor of 2  over the corresponding condensed (ordered) term. Such a term will replace the full population term $N_{i}$ whenever the index $i$ is multiplied by a condensate amplitude $z_{i}$. By analogy, $N_{i} \rightarrow N_{i}^{TT}$ whenever the population term appears multiplied by two condensate mean amplitudes $z_{i} z_{i}$ (i.e. condensate anomalous averages of the form $zz$, or $z^{*}z^{*}$, but {\em not} normal averages $z^{*}z$).

In order to keep the notation general, so that it can be easily applicable to n partially condensed levels with next neighbour interactions, we have also defined the following rates 
\begin{equation}
\nu_{(i \mp 1)i}^{(i \pm 1)} = I_{0} \left[ 4V_{(i \mp 1) i i (i \mp 1)}-
 2V_{(i \mp 1)(i \pm 1)(i \pm 1)(i \mp 1)}- V_{(i \mp 1)(i \mp 1)(i \mp 1)(i \mp 1)} \right] V_{(i \mp 1)(i \pm 1) i i} \label{nu}
\end{equation} 
\begin{equation}
\vartheta_{(i \mp 1)i}^{(i \pm 1)} = I_{0} \left[ V_{(i \mp 1) i i (i \mp 1)} + V_{i (i \pm 1)(i \pm 1)i}- V_{iiii} \right] V_{(i \mp 1)(i \pm 1) i i} \label{theta}
\end{equation}
\begin{equation}
\varepsilon_{(i \mp 1)i}^{(i \pm 1)} = I_{0} \left[ 2V_{(i \mp 1)(i \pm 1)(i \pm 1)(i \mp 1)}- V_{iiii} \right] V_{(i \mp 1)(i \pm 1) i i} \label{eps}
\end{equation}
\begin{equation}
\lambda_{ijji}^{kllk} =  V_{ijji} V_{kllk} \lim_{\eta \rightarrow 0^{+}}  \int_{0}^{\infty} d \tau e^{- \eta \tau }  
\end{equation}
\begin{equation}
\gamma_{ijji}^{0211} =  I_{0} \left( V_{ijji} V_{0211} \right)
\end{equation}
where
\begin{equation}
I_{0} = \frac{1}{2} \left\{ \lim_{\eta \rightarrow 0^{+}}  \int_{0}^{\infty} d \tau e^{ \pm i(\omega_{0}+\omega_{2}-2\omega_{1} \pm i \eta) \tau } +   \lim_{\eta \rightarrow 0^{+}} \int_{0}^{\infty} d \tau e^{- \eta \tau } \right\}
\end{equation}

For the condensed components of the two lowest states, we obtain the following off-diagonal contributions
\begin{eqnarray}
\tilde{X}_{00}(V^{2}) =  & &   - 4 |z_{0}|^{2} \left\{ \begin{array}{l} 2 \lambda_{0000}^{0110} \rho_{01} \rho_{10} + \left( \Gamma_{0211} + 2 \lambda_{0110}^{0220} \right) \rho_{12} \rho_{21} + 2\lambda_{0000}^{0220} \rho_{02} \rho_{20} \\ +2 \left( 2 \gamma_{0110}^{0211} - \gamma_{0220}^{0211} \right) \rho_{10} \rho_{12} \end{array} \right\} \nonumber \\
& & +4 |z_{1}|^{2} \left[ \lambda_{0110} \rho_{01} \rho_{10} + \Gamma_{0211} \rho_{12} \rho_{21} + 2\gamma_{0110}^{0211} \rho_{10} \rho_{12} \right]  \nonumber \\
& & +\zeta_{10} \left\{ \begin{array}{l} -2 |z_{0}|^{2} \left[ 4\lambda_{0000}^{0110} \rho_{01} + \left( 4 \gamma_{0110}^{0211}-2 \gamma_{0220}^{0211} - \gamma_{0000}^{0211} \right) \rho_{12} \right] \\
 +2 |z_{1}|^{2} \left[ \Gamma_{0211} \rho_{01} + \left( \gamma_{0110}^{0211}+3 \gamma_{1221}^{0211} -3 \gamma_{1111}^{0211} \right) \rho_{12} \right] \\
-2 \left[ 2 \lambda_{0110}  \rho_{01} + \left( 2 \gamma_{0220}^{0211}+\gamma_{1111}^{0211} \right) \rho_{12} \right] \\
+4 \rho_{01} \left[ \Gamma_{0211} (n_{1}-n_{2}) - 2 \lambda_{0110} n_{1} \right] \\
-4 \rho_{12} \left[ -\gamma_{0000}^{0211} n_{0} + \left( \gamma_{0110}^{0211} +3 \gamma_{1111}^{0211}-3\gamma_{1221}^{0211} \right) n_{1} +   \left( 4\gamma_{1221}^{0211} - \gamma_{2222}^{0211} \right) n_{2} \right] \\
-4 \left[ \left( \lambda_{0211} + 2 \lambda_{0110}^{0220} \right) \rho_{02} \rho_{21} + \gamma_{0000}^{0211} \rho_{10} \rho_{02} \right] \\
-2 \gamma_{0000}^{0211} \zeta_{10} \rho_{02} \end{array} \right\} \label{X0}
\end{eqnarray}

\begin{eqnarray}
\tilde{X}_{11}(V^{2}) = & & - 4 |z_{1}|^{2} \left\{ \begin{array}{l} \left( \Gamma_{0211}+2 \lambda_{1111}^{0110} \right) \rho_{01} \rho_{10} + \left( \Gamma_{0211} + 2 \lambda_{1111}^{1221} \right) \rho_{12} \rho_{21} + \left( -\Gamma_{0211} + 2\lambda_{0110}^{1221} \right) \rho_{02} \rho_{20} \\ +2 \left( 2 \gamma_{0110}^{0211} +2 \gamma_{1221}^{0211} - \gamma_{1111}^{0211} \right) \rho_{10} \rho_{12} \end{array} \right\} \nonumber \\
& & +4 |z_{0}|^{2} \left[ \lambda_{0110} \rho_{01} \rho_{10} + \Gamma_{0211} \rho_{12} \rho_{21} + 2\gamma_{0110}^{0211} \rho_{10} \rho_{12} \right]  \nonumber \\
& & +\zeta_{10} \left\{ \begin{array}{l} -2 |z_{1}|^{2} \left[ \left( 3 \Gamma_{0211} + 4\lambda_{1111}^{0110} \right) \rho_{01} + \left( 5 \gamma_{0110}^{0211} +5 \gamma_{1221}^{0211} - 3 \gamma_{1111}^{0211} \right) \rho_{12} \right] \\
 +2 |z_{0}|^{2} \left[ \hspace{1.0cm}  \left( 4 \gamma_{0110}^{0211} -2 \gamma_{0220}^{0211} - \gamma_{0000}^{0211} \right) \rho_{12} \right] \\
-2 \left[ 2 \lambda_{0110}  \rho_{01} + \left( 2 \gamma_{0220}^{0211}+\gamma_{1111}^{0211} \right) \rho_{12} \right] \\
+4 \rho_{01} \left[ - \Gamma_{0211} (n_{1}-n_{2}) - 2 \lambda_{0110} n_{0} \right] \\
-4 \rho_{12} \left[ \begin{array}{l} \left( \gamma_{0000}^{0211} +2 \gamma_{0220}^{0211}-2\gamma_{0110}^{0211} \right)  n_{0} + \left( \gamma_{0110}^{0211} + \gamma_{1221}^{0211}-\gamma_{1111}^{0211} \right) n_{1} \\ +   \left( \gamma_{2222}^{0211} +2 \gamma_{0220}^{0211} -2 \gamma_{1221}^{0211} \right) n_{2} \end{array} \right] \\
-4 \left[ \left( -\Gamma_{0211} + 2 \lambda_{0110}^{1221} \right) \rho_{02} \rho_{21} + \left( \gamma_{0000}^{0211} +2 \gamma_{0220}^{0211} \right) \rho_{10} \rho_{02} \right] \\
-2 \left( \gamma_{0000}^{0211} + 2 \gamma_{0220}^{0211} \right) \zeta_{10} \rho_{02} \end{array} \right\} \label{X1}
\end{eqnarray}

Although it is exteremely hard to discuss the physical implication of each separate contribution appearing in the above equations, at this point we would like to comment briefly on a carefully selected subset of the above equations, namely
\begin{equation}
\frac{d |z_{0}|^{2}}{dt} = 8 |z_{1}|^{2} \left\{ \lambda_{0110} \rho_{01} \rho_{10} +  \Gamma_{0211} \rho_{12} \rho_{21} + 2  \gamma_{0110}^{0211} \rho_{10} \rho_{12} \right]
\end{equation}
\begin{equation}
\frac{d |z_{1}|^{2}}{dt} = 8 |z_{0}|^{2} \left\{ \lambda_{0110} \rho_{01} \rho_{10} +  \Gamma_{0211} \rho_{12} \rho_{21} + 2  \gamma_{0110}^{0211} \rho_{10} \rho_{12} \right]
\end{equation}
Such terms indicate clearly a mechanism of `coherent population transfer', that is growth of condensation in one (bare) level due to the existence of condensation in another (bare) level, which  occurs via off-diagonal incoherent couplings $\rho_{ij}$. By further ignoring, for simplicity, the redistributional processes ($V_{0211} \rightarrow 0$) we see a sort of Rabi-like oscillation between the two condensates, via the process
\begin{equation}
\frac{d |z_{i}|^{2} }{dt} \sim |V_{ijji}|^{2} \left[ \rho_{ij} \rho_{ji} \right] |z_{j}|^{2} \label{Jos}
\end{equation}
Such processes do not lead to changes in the total trap level populations. On the contrary, due to the co-existence of coherent and incoherent atoms within the same trap level, this term can be interpreted as follows: If one level initially contains a non-vanishing coherent amplitude, this will be gradually transferred to the other level (even if the other level is initially fully incoherent) without an associated change in the total population of each level (since the incoherent population $n_{i}$ of each level adjusts accordingly to keep $N_{i}$ fixed). Hence, even if we assume that the condensate initially resides only on the bottom trap level, the above contribution generated by nonlinear coupling interactions will tend to give rise to a coherent mean field amplitude in level 1 (and vice versa). Whether this term actually becomes important at all (and in what limits this may be so) depends on how heavily this contribution is overshadowed by all other existing terms in the equations of motion for coherent evolution. Although we would not expect this term to play a significant role in the full population dynamics, this is something which should be confirmed by direct numerical simulation of the full equations given in this paper. We conclude the discussion by noting that such processes may become important in the case of coupled 2-species condensation \cite{HF_Theory,HF_Exp}, or in the creation of non-ground state trapped condensates \cite{Non-Ground_BEC}, whereby the population between such levels is controlled by the application of external fields.

The off-diagonal contributions to the uncondensed dynamics of the two lowest levels are given by
\begin{eqnarray}
Y_{00}(V^{2})  = & &  -2 \Gamma_{0211} \left\{ \rho_{02} \rho_{20} +2   \left[ \rho_{12} \rho_{21} n_{0} - \left( \rho_{01} \rho_{10} + \rho_{12} \rho_{21} - \rho_{02} \rho_{20} \right) n_{1} + \rho_{01} \rho_{10} n_{2} \right] \right\} \nonumber \\
& & -2  \left[ 4 \Gamma_{0211} \rho_{02} \rho_{21} \rho_{10} + \nu_{01}^{(2)} \rho_{10}^{2} \rho_{02} + \nu_{21}^{(0)} \rho_{12}^{2} \rho_{20} \right] \nonumber \\
& & +8 |z_{0}|^{2} \left[ \lambda_{0000}^{0110} \rho_{01} \rho_{10} + \lambda_{0110}^{0220} \rho_{12} \rho_{21} + \lambda_{0000}^{0220} \rho_{02} \rho_{20} \right]  \nonumber \\
& & +4 |z_{1}|^{2} \left[ \left(\Gamma_{0211}- \lambda_{0110} \right) \rho_{01} \rho_{10} -\Gamma_{0211} \rho_{02} \rho_{20} \right]   \nonumber \\
& & + \rho_{10} \rho_{12} \left\{ \begin{array}{l} 2 \left( \gamma_{0220}^{0211}- \gamma_{1111}^{0211} \right) -4  \left[ \nu_{01}^{(2)} n_{0} - 3 \vartheta_{01}^{(2)} n_{1} + \nu_{21}^{(0)} n_{2} \right]  \\
+4 |z_{0}|^{2} \gamma_{0000}^{0211} +4 |z_{1}|^{2} \left( \gamma_{0110}^{0211} + 3 \gamma_{1221}^{0211} - 3 \gamma_{1111}^{0211} \right) \end{array}  \right\}  \nonumber \\
& & +\zeta_{10} \left\{ \begin{array}{l} 8 \lambda_{0000}^{0110} |z_{0}|^{2} \rho_{01}  \\
 +2 |z_{1}|^{2} \left[ \Gamma_{0211} \rho_{01} + 2 \gamma_{0110}^{0211} \rho_{12} \right] \\
+4 \lambda_{0110} \rho_{01} +8 \gamma_{0220}^{0211} \rho_{12} \\
+4 \rho_{01} \left[ \left( 2\lambda_{0110} + \Gamma_{0211}\right) n_{1} - \Gamma_{0211} n_{2} \right] \\
+4 \rho_{12} \left[ 2 \left(  \gamma_{0220}^{0211} -2 \gamma_{0110}^{0211} \right) n_{0} + 4 \gamma_{0110}^{0211} n_{1} +2\gamma_{0220}^{0211} n_{2} \right] \\
+4 \rho_{02} \rho_{21} \left(2 \lambda_{0110}^{0220}- \Gamma_{0211} \right) +8 \rho_{10} \rho_{02} \left( \gamma_{0000}^{0211} -2 \gamma_{0110}^{0211} + \gamma_{0220}^{0211} \right) \\
+ \zeta_{10} \rho_{02} \left( 3\gamma_{0000}^{0211} -4\gamma_{0110}^{0211} +2\gamma_{0220}^{0211} \right) \end{array} \right\} \label{Y0}
\end{eqnarray}
and
\begin{eqnarray}
Y_{11}(V^{2}) =  & & 4 \Gamma_{0211} \left\{ \rho_{02} \rho_{20} +2 \left[ \rho_{12} \rho_{21} n_{0} - \left( \rho_{01} \rho_{10} + \rho_{12} \rho_{21} - \rho_{02} \rho_{20} \right) n_{1} + \rho_{01} \rho_{10} n_{2} \right] \right\} \nonumber \\
& & +4   \left[ 4 \Gamma_{0211} \rho_{02} \rho_{21} \rho_{10} + \nu_{01}^{(2)} \rho_{10}^{2} \rho_{02} + \nu_{21}^{(0)} \rho_{12}^{2} \rho_{20} \right] \nonumber \\
& & +4 |z_{0}|^{2} \left[ - \lambda_{0110} \rho_{01} \rho_{10} + \Gamma_{0211} \rho_{12} \rho_{21} \right]  \nonumber \\
& & +4 |z_{1}|^{2} \left[ \left(2 \lambda_{0110}^{1111} - \Gamma_{0211} \right) \rho_{01} \rho_{10} + \left(2 \lambda_{1111}^{1221} -\Gamma_{0211} \right) \rho_{12} \rho_{21} + \left( 2\lambda_{0110}^{1221} + \Gamma_{0211}  \right) \rho_{02} \rho_{20} \right]  \nonumber \\
& & + \rho_{10} \rho_{12} \left\{ \begin{array}{l} -4 \left( 2\gamma_{0220}^{0211}- \gamma_{1111}^{0211} \right) +8   \left[ \nu_{01}^{(2)} n_{0} - 3 \vartheta_{01}^{(2)} n_{1} + \nu_{21}^{(0)} n_{2} \right] \\
+8 |z_{0}|^{2} \left( 3\gamma_{0110}^{0211} -2 \gamma_{0220}^{0211} -  \gamma_{0000}^{0211} \right) +8 |z_{1}|^{2} \left( 2\gamma_{1111}^{0211} - \gamma_{0110}^{0211} -  \gamma_{1221}^{0211} \right) \end{array}  \right\}  \nonumber \\
& & +\zeta_{10} \left\{ \begin{array}{l} 2 \nu_{01}^{(2)}  |z_{0}|^{2} \rho_{12}  \\
 +2 |z_{1}|^{2} \left[ \left( 4 \lambda_{0110}^{1111} - \Gamma_{0211} \right) \rho_{01} + \left( 3 \gamma_{1111}^{0211} - \gamma_{0000}^{0211} - \gamma_{1221}^{0211} \right) \rho_{12} \right] \\
+4 \lambda_{0110} \rho_{01} + 2 \left( 3 \gamma_{1111}^{0211} - 2 \gamma_{0220}^{0211} \right) \rho_{12} \\
+4 \rho_{01} \left[ 2 \lambda_{0110} n_{0} - 3 \Gamma_{0211} n_{1} + 3 \Gamma_{0211} n_{2} \right]  \\
+4 \rho_{12} \left[ \left( 6 \gamma_{0110}^{0211} -2 \gamma_{0220}^{0211} - \gamma_{0000}^{0211} \right) n_{0} -5 \vartheta_{01}^{(2)} n_{1} +  \left( 6 \gamma_{1221}^{0211} -2 \gamma_{0220}^{0211} - \gamma_{2222}^{0211} \right) n_{2} \right] \\
+4 \rho_{02} \rho_{21} \left(2 \lambda_{0110}^{1221}+3 \Gamma_{0211} \right) +4 \rho_{10} \rho_{02} \left( 8 \gamma_{0110}^{0211} -2 \gamma_{0220}^{0211} - \gamma_{0000}^{0211} \right) \\
+8 \zeta_{10} \rho_{02} \gamma_{0110}^{0211}  \end{array} \right\}
\end{eqnarray}

\section{Derivation of Condensate Evolution}

The second order contribution to $|z_{i}|^{2}$ of Eq. (17) can be found by adiabatically eliminating the quantity $z_{i}^{*} \langle \hat{a}_{s}^{\dag} \hat{a}_{m} \hat{a}_{r} \rangle$ via
\begin{equation}
i \frac{d z_{i}}{dt} = \omega_{i} z_{i} + \sum_{pql}^{'} T_{ipql} \langle \hat{a}_{p}^{\dag} \hat{a}_{q} \hat{a}_{l} \rangle \label{app}
\end{equation}
and
\begin{equation}
i  \frac{d}{dt}\langle \hat{a}_{s}^{\dag}\hat{a}_{m}\hat{a}_{r}  \rangle  = (\omega_{m}+\omega_{r}-\omega_{s}) \langle \hat{a}_{s}^{\dag}\hat{a}_{m}\hat{a}_{r}  \rangle  + f_{\langle \hat{a}^{\dag} \hat{a} \hat{a} \rangle}
\end{equation}
with
\begin{eqnarray}
f_{\langle \hat{a}^{\dag} \hat{a} \hat{a} \rangle} & & =  \sum_{lt} T_{mrlt} \langle \hat{a}_{s}^{\dag}\hat{a}_{l}\hat{a}_{t}  \rangle  \nonumber \\
& & +  \sum_{plt} T_{prlt} \langle \hat{a}_{p}^{\dag} \hat{a}_{s}^{\dag} \hat{a}_{m}\hat{a}_{l} \hat{a}_{t}  \rangle + \sum_{plt} T_{pmlt} \langle \hat{a}_{p}^{\dag} \hat{a}_{s}^{\dag} \hat{a}_{r} \hat{a}_{l} \hat{a}_{t}  \rangle -  \sum_{pql} T_{pqls} \langle \hat{a}_{p}^{\dag} \hat{a}_{q}^{\dag} \hat{a}_{l} \hat{a}_{m} \hat{a}_{r} \rangle 
\end{eqnarray}
Hence, the quantity $F_{\langle \hat{a}^{\dag} \hat{a} \hat{a} \rangle}$ corresponding to $F$ of Eq. (\ref{F}), now becomes
\begin{equation}
F_{\langle \hat{a}^{\dag} \hat{a}\hat{a}  \rangle} = z_{i}^{*} f_{\langle \hat{a}^{\dag} \hat{a}\hat{a}  \rangle} - \sum_{pql} T_{pqli} \langle \hat{a}_{p}^{\dag}\hat{a}_{q}^{\dag} \hat{a}_{l}  \rangle \langle \hat{a}_{s}^{\dag}\hat{a}_{m}\hat{a}_{r}  \rangle
\end{equation}

\section{Second Order Collisional Integrals}

For completeness, we give here the second order collisional integrals of Walser et al. \cite{Walser_QK} as generated by our approach.

\begin{eqnarray}
\left( \frac{d \rho}{dt} \right)^{Walser} & = & 2 \sum_{rsmn} \sum_{pqlt} T_{rsmn} T_{pqlt}^{(\delta)} \nonumber \\
& \times & \left\{ \begin{array}{l}  \left[ \left( \rho_{mp} + \delta_{mp} \right) \left( \rho_{nq} + \delta_{nq} \right) \rho_{ts} \rho_{lr} - \rho_{mp} \rho_{nq} \left( \rho_{ts} + \delta_{ts} \right)   \left( \rho_{lr} + \delta_{lr} \right) \right] \\ 
+ 2  \left[ \left( \rho_{mp} + \delta_{mp} \right) \left( z_{q}^{*} z_{n} \right) \rho_{ts} \rho_{lr} - \rho_{mp} \left( z_{q}^{*} z_{n} \right) \left( \rho_{ts} + \delta_{ts} \right)  \left( \rho_{lr} + \delta_{lr} \right) \right] \\ 
+  \left[ \left( \rho_{mp} + \delta_{mp} \right) \left( \rho_{nq} + \delta_{nq} \right) \left( z_{s}^{*} z_{t} \right) \rho_{lr} - \rho_{mp} \rho_{rq} \left( z_{s}^{*} z_{t} \right) \left( \rho_{lr} + \delta_{lr} \right) \right] \\  
+ 2 \left[ \left( \rho_{mp} + \delta_{mp} \right) \kappa_{nt} \kappa_{qs}^{*} \rho_{lr} - \rho_{mp} \kappa_{nt} \kappa_{qs}^{*}  \left( \rho_{lr} + \delta_{lr} \right)  \right] \\  
+ 2 \left[ \left( \rho_{mp} + \delta_{mp} \right) \kappa_{nt} \kappa_{qs}^{*}  \left( z_{r}^{*} z_{l} \right) - \rho_{mp} \kappa_{nt} \kappa_{qs}^{*}  \left( z_{r}^{*} z_{l} \right)  \right] \\  
+ 2 \left[ \left( z_{p}^{*} z_{m} \right) \kappa_{nt} \kappa_{qs}^{*} \rho_{lr} -  \left( z_{p}^{*} z_{m} \right)  \kappa_{rt} \kappa_{qs}^{*} \left( \rho_{lr} + \delta_{lr} \right)  \right] \\ 
+ 2 \left[ \left( \rho_{mp} + \delta_{mp} \right) \left( z_{n} z_{t} \right) \kappa_{qs}^{*} \rho_{lr} - \rho_{mp}  \left( z_{n} z_{t} \right)  \kappa_{qs}^{*} \left( \rho_{lr} + \delta_{lr} \right) \right] \\   
+4 \left[ \left( \rho_{mp} + \delta_{mp} \right) \kappa_{nt} \left( z_{q}^{*} z_{s}^{*} \right) \rho_{lr} - \rho_{mp} \kappa_{nt}  \left( z_{q}^{*} z_{s}^{*} \right)  \left( \rho_{lr} + \delta_{lr} \right)  \right] \\ 
+ 2 \kappa_{qi}^{*} \left[ \begin{array}{l} \left[ \left( \rho_{mp} + \delta_{mp} \right) \kappa_{rl}  \rho_{ts} - \rho_{mp} \kappa_{rl} \left( \rho_{ts} + \delta_{ts} \right)  \right] \\ 
+ \left[ \left( z_{p}^{*} z_{m} \right) \kappa_{rl}  \rho_{ts} -  \left( z_{p}^{*} z_{m} \right)   \kappa_{rl} \left( \rho_{ts} + \delta_{ts} \right)  \right] \\
+ \left[ \left( \rho_{mp} + \delta_{mp} \right) \kappa_{rl} \left( z_{s}^{*} z_{t} \right)  - \rho_{mp} \kappa_{rl} \left( z_{s}^{*} z_{t} \right) \right] \\
+ \left[ \left( \rho_{mp} + \delta_{mp} \right) \left( z_{r} z_{l} \right) \rho_{ts} - \rho_{mp}  \left( z_{r} z_{l} \right)  \left( \rho_{ts} + \delta_{ts} \right)  \right] \end{array} \right]
\end{array} \right\}
\end{eqnarray}
As explained in the text
\begin{equation}
T_{pqlt}^{(\delta)} = \int dt^{'} e^{- i \left( \omega_{l} +  \omega_{t} -  \omega_{p} -  \omega_{q} \right) (t - t^{'}) } = \pi \delta(\Delta \omega) - i P \left( \frac{1}{ \Delta \omega} \right)
\end{equation}
Here $ \Delta \omega = \left( \omega_{l} + \omega_{t} - \omega_{p} - \omega_{q} \right)$ and the $ \omega_{i}$ denotes the eigenenergy of level $i$ in the particular basis chosen for the analysis of the system. In our original formulation, these correspond to bare trap eigenenergies, whereas in order to establish exact analogy with the collisional integrals of Walser et al. \cite{Walser_QK}, these should be replaced by their corresponding values dressed by normal {\em and} anomalous HFB mean fields. 

In the original formulation of Walser et al., their basis is left unspecified, and their resulting expressions are assumed to be valid for any single-particle basis dressed by, at most, {\em number-conserving} mean fields (i.e. no anomalous averages), a conclusion {\em not} supported by our analysis.


\begin{thebibliography}{99}
\bibitem{Comment} The reason for this is that the many-body T-matrix used here \cite{Stoof_NIST} is not the fully-dressed (quasiparticle) T-matrix for binary collisions (found in \cite{Bijlsma}), but it includes only bare diagrams and thus corresponds, in the homogeneous limit, to the zero-energy, zero-momentum limit of the many-body T-matrix \cite{Keith,Prouk_1D}; in this limit, such `bubble' terms are ignored, so that a consistent formulation in terms of $T_{ijji}^{MB}$ should not explicitly include any contributions of the form $(n_{i}-n_{j})$.
\end{thebibliography}
\end{document}